\documentclass[sigconf]{acmart}
\AtBeginDocument{%
  \providecommand\BibTeX{{%
    \normalfont B\kern-0.5em{\scshape i\kern-0.25em b}\kern-0.8em\TeX}}}

\setcopyright{acmcopyright}
\copyrightyear{2023}
\acmYear{2023}
\acmDOI{10.1145/3544548.3580734}
\acmConference[CHI '23]{Proceedings of the 2023 CHI Conference on Human Factors in Computing Systems}{April 23--28, 2023}{Hamburg, Germany}
\acmBooktitle{Proceedings of the 2023 CHI Conference on Human Factors in Computing Systems (CHI '23), April 23--28, 2023, Hamburg, Germany}
\acmPrice{15.00}
\acmISBN{978-1-4503-XXXX-X/18/06}

\usepackage{soul}
\usepackage{enumitem}
\usepackage{wrapfig}
\usepackage{multirow}
\usepackage{tabu}                      
\usepackage{booktabs}                  
\usepackage{graphicx}
\usepackage{algorithm}
\usepackage{algpseudocode}
\usepackage{makecell}

\newcommand{\eg}{e.g. }

\definecolor{red}{rgb}{1,0,0}

\definecolor{marked}{rgb}{0,0,0}
\newcommand{\yh}[1]{{\color{marked}#1}}

\newcommand{\od}[1]{{\color{marked}#1}}

\newcommand{\revised}[1]{{\color{marked}#1}}
\newcommand{\rereview}[1]{{\color[rgb]{0,0,0} #1}}

\hypersetup{
    colorlinks=false,
    pdfborder={0 0 0},
    pdfborderstyle={/S/U/W 0},
}

\begin{document}

\title{{Interactive Context-Preserving Color Highlighting} for\\ Multiclass Scatterplots}


\author{Kecheng Lu}
\email{lukecheng0407@gmail.com}
\orcid{0000-0001-5990-3296}
\affiliation{%
  \institution{Shandong University}
  \country{China}
}

\author{Khairi Reda}
\email{redak@iu.edu}
\orcid{0000-0002-8096-658X}
\affiliation{%
  \institution{Indiana University-Purdue University Indianapolis}
  \country{United States}
}

\author{Oliver Deussen}
\email{oliver.deussen@uni-konstanz.de}
\orcid{0000-0001-5803-2185}
\affiliation{%
  \institution{University of Konstanz}
  \country{Germany}
}

\author{Yunhai Wang}
\authornote{corresponding author}
\email{cloudseawang@gmail.com}
\orcid{0000-0003-0059-6580}
\affiliation{%
  \institution{Shandong University}
  \country{China}
}
\renewcommand{\shortauthors}{Lu et al.}


\begin{abstract}

Color is one of the main visual channels used for highlighting elements of interest in visualization. However, in multi-class scatterplots, color highlighting often comes at the expense of degraded color discriminability. In this paper, we argue for \yh{\emph{context-preserving highlighting} during the} interactive exploration of multi-class scatterplots to achieve desired pop-out effects, while maintaining good perceptual separability among all classes and consistent color mapping schemes under varying \yh{points} of interest. We do this by first generating two contrastive color mapping schemes with large and small contrasts to the background. \yh{Both schemes maintain good perceptual separability among all classes and ensure that when colors from the two palettes are assigned to the same class, they have a high color consistency in color names.} We then interactively combine these two schemes to create a dynamic color mapping for highlighting different \yh{points} of interest. We demonstrate the effectiveness through crowd-sourced experiments and case studies.


\end{abstract}



\begin{teaserfigure}\vspace*{-12pt}
  \includegraphics[width=.95\textwidth]{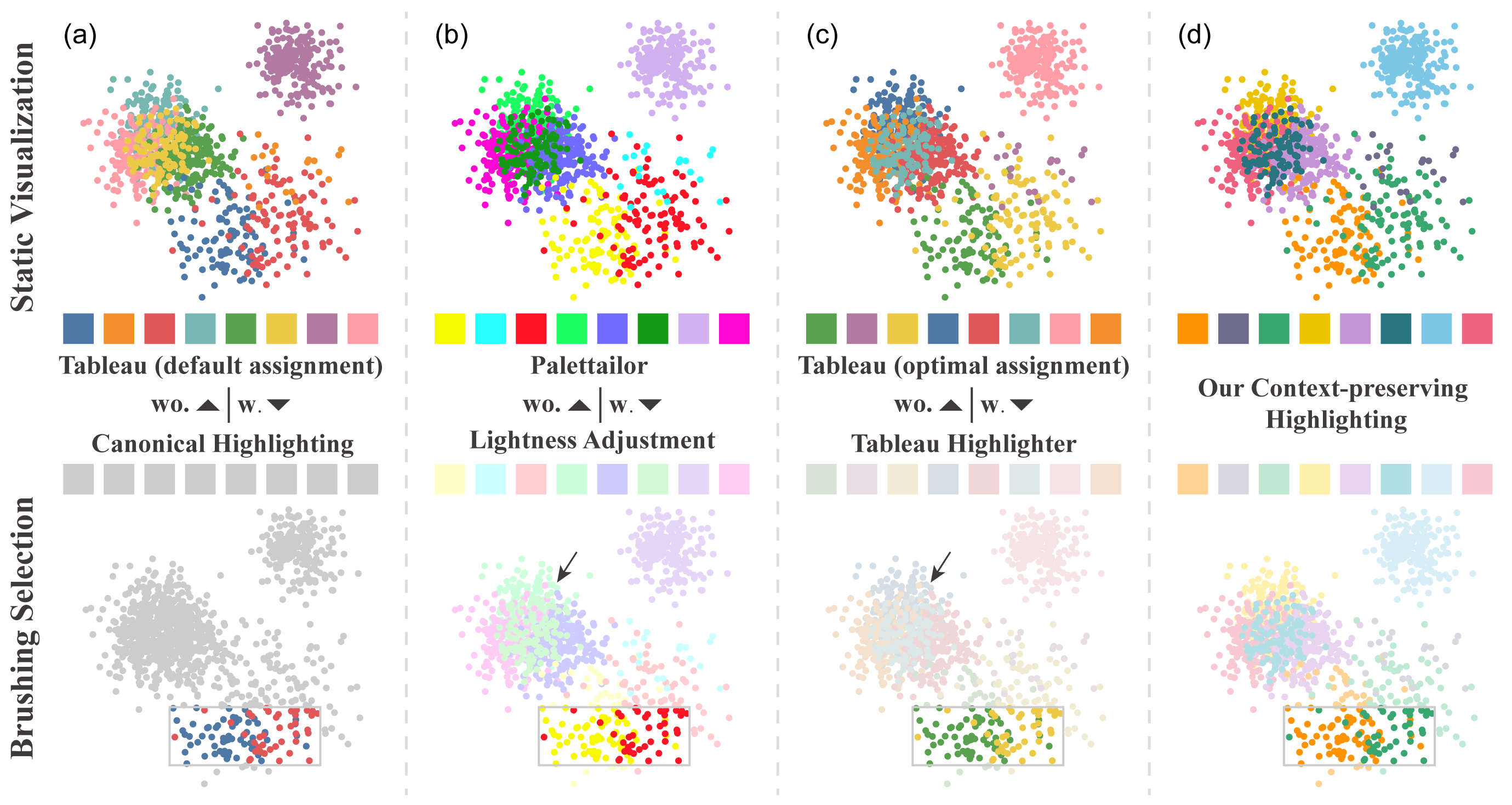}
  \caption{\od{Results for applying different color-based highlighting methods for highlighting points in
  a multi-class scatterplot. 
  \textbf{(a)} (top) colorization with
  Tableau palette and default assignment; (bottom)
    highlighting effect achieved by assigning a grey color to all non-selected data points;
  \textbf{(b)} (top) result for a
  Palettailor-generated palette~\cite{Lu21};  (bottom) highlighting achieved by increasing lightness of non-selected data points;
    \textbf{(c)} (top) colorization by Tableau palette and optimal assignment; (bottom) highlighting by applying Tableau Highlighter function;
     \textbf{(d)} (top) colorization by our method using a salient color palette; (bottom) highlighting result by combining salient and faint color mapping schemes.
     \rereview{Problematic areas are indicated by arrows.}
     Our method allows highlighting a subset of data points while maintaining the discriminability of all non-selected points as well as the color consistency of all color pairs.}}
  \label{fig:teaser}\vspace*{-4pt}
\end{teaserfigure}

\maketitle

\section{Introduction}
Multi-class scatterplots are among the most commonly used representations for visualizing labeled quantitative data. They represent each data item with a color-coded point (or some other marks)
positioned within two orthogonal dimensions, with color encoding class label (i.e., category).
If used with judiciously designed color palettes, multi-class scatterplots can effectively display the distribution of classes and the relationships among them. A few automated colorization methods~\cite{Wang2018, Lu21} have been proposed for maximizing class discriminability in scatterplots while maintaining their aesthetic appeal. \yh{
Typically, all classes in a visualization are given equal emphasis (see the first row in Fig.~\ref{fig:teaser}). This type of color assignment is sufficient for a static visualization.} 

\yh{For interactive visualizations, however, it is desirable to emphasize certain subsets of the data on-demand, for example, to allow the user to select or brush points of interest. The canonical strategy to support this interaction is to dim colors for the non-selected data by assigning a neutral grey. Although effective at inducing a highlight, this approach results in a loss of context (\eg see Fig.~\ref{fig:teaser} (a)-bottom).
While modulating lightness or opacity for the non-selected data items can alleviate this issue~\cite{robinson2011highlighting}, it can lead to poor class separability (\eg  see the small difference between the green and light-green classes in the bottom of Fig.~\ref{fig:teaser} (b)). In short, existing color-based highlighting strategies cause a temporary loss of color coding~\cite{munzner2014visualization}, \od{which disrupts exploration or, at minimum, the user's mental map.} Even state-of-the-art commercial solutions, such as Tableau Highlighter~\cite{tableauhighlighter}, struggle at producing good color highlighting while preserving context (see Fig.~\ref{fig:teaser} (c)-bottom).}
Designing color palettes that support both, \emph{focus and context}, is therefore crucial for interactive visualization. Yet, this important design goal is unsupported by existing colorization approaches (e.g., Palettailor~\cite{Lu21} and Colorgorical~\cite{Gramazio17}), as those approaches do not allow for \od{dynamically varying the visual emphasis of selected data points}.

We present an interactive colorization technique for the \revised{\emph{context-preserving} highlighting} of scatterplots (and for other multi-class labeled visualizations).  The core of this technique is the ability to interactively emphasize data points of interest while also maintaining a good perceptual separability for all classes, and furthermore ensuring color consistency under varying degrees of emphasis.
Our technique allows the user to
highlight elements in a visualization while still being able to see the local neighborhood and global context of those elements. \revised{This allows for creating compelling visualizations, such as multi-view interfaces that support effective brushing-and-linking without loss of context (see supplementary video).}

The proposed technique works by generating two contrastive color mapping schemes that maximize and minimize the contrast over the background, while
both optimize the discriminability for a given scatterplot configuration.
We create the needed emphasis by interactively combining both palettes. To achieve effective highlighting, we model the contrast to the background as well as the class-neighborhood contrast. Additionally, our technique maintains color consistency for  hue, saturation, and color names between highlighted and de-emphasized colors for each class, thus ensuring minimal disruption to the user's mental model. Fig.~\ref{fig:teaser} (d) shows two pre-generated palettes of our method. The bottom of (d) shows the results of interactive highlighting. 
For the purpose of letting the selected data points of interest stand out, our results have a similar effect to Figs.~\ref{fig:teaser} (a, b, c). However, in contrast to other methods, our method ensures good class separability and color names among selected and non-selected data points.

We evaluated our approach in two crowd-sourced studies\footnote{Experimental data and analysis code are included with the submission as supplemental materials \rereview{and are available at \url{https://osf.io/679pb/}}.}. First, we compared our colorization results with state-of-the-art palettes  (e.g., Tableau~\cite{tableau} and Palettailor~\cite{Lu21}). The results indicate that our method achieves a comparable highlighting effect while better maintaining class separability than the benchmark methods.
We also created a web-based implementation of our technique as a color-design tool\footnote{ \url{https://palettailor.github.io/highlighting/}}
and demonstrate its effectiveness in a case study. To summarize, the main contributions of this paper are as follows:

\vspace*{-1mm}
\begin{itemize}[noitemsep]
\setlength{\itemsep}{5pt}
  \item We propose an interactive context-\yh{preserving}  approach for generating stable color palettes for multi-class visualizations. Our approach supports an interactive emphasis on data parts while maintaining overall class discriminability and relative color consistency.
  \item We present a simulated annealing-based optimization for generating highlightable palettes, while ensuring sufficient contrast with the background and neighboring classes, among other perceptual constraints.
   \item   We empirically validate our techniques in two crowd-sourced experiments and present extensions of our method to \yh{a few interactions and other multi-class visualizations such as line and bar charts, within an open-source implementation\footnote{ \url{https://palettailor.github.io/highlighting/demo/}}.} 
\end{itemize}

\section {Related Work}
We divide previous works into methods related to visual highlighting and to color design for visualization.
\yh{\subsection{Highlighting in Interactive Visualization}
\od{In interactive visualization applications, it is a common task to highlight a subset of data points 
for directing the user attention or apply subsequent operations to}~\cite{liang2010highlighting,robinson2011highlighting,strobelt2015guidelines}.}
Emphasis effects are created by manipulating visual variables (\eg position, size, transparency and color lightness). Naidu~\cite{naidu2019measuring} presents a crowd-sourced study to measure the highlighting effect of color-coded scatterplots and provides recommendations for effective color highlighting.

Previous studies have systematically evaluated emphasis effects in a wider range of
scenarios~\cite{griffin2014comparing,waldner2017exploring,mairena2021emphasis}.
Hall et al.~\cite{hall2016formalizing} provide a systematic review of such effects and divide them into two classes: intrinsic and extrinsic effects. The former is created by the initial visual mapping, while the latter is the result of manipulating the visual variables of an existing visualization.
Although the extrinsic emphasis is effective in many cases, it may conflict with the
visual encoding of the given visualization. For example, changing transparency or color lightness in a multi-class scatterplot may result in new colors (see Fig.~\ref{fig:teaser} (b)) that might lead to misunderstanding color-associated semantics or to similar colors (see Fig.~\ref{fig:teaser} (c)) that do not allow visual discrimination anymore.
To address this issue, our approach generates stable color mapping schemes for interactively highlighting multi-class scatterplots that attempt to balance between two goals: emphasizing \revised{points of interest} \emph{and} maintaining class discrimination and \yh{color consistency}.

\subsection{Color Design in Visualization}
For a complete review of color design techniques for visualization, we refer readers to surveys such as~\cite{Tominski08, Zhou16}. We limit our discussion to techniques related to color design for categorical data visualization and specifically to the optimization of color mappings, color palette generation, color palettes for highlighting  as well as \yh{color consistency}.

\vspace{1.5mm}
\noindent\textbf{Color Map Optimization}. Mapping each class to a proper color selected from a given palette is particularly helpful for categorical data visualization since no given order can be used here.
A few factors have been identified for guiding searches within such mappings.
For example, Lin et al.~\cite{lin2013selecting} propose to optimize the compatibility between class semantics and assigned colors. Setlur and Stone~\cite{setlur2016linguistic} produce better results by using co-occurrence measures of color name frequencies. Reda et al. argue generally for increasing the nameability of colors in colormaps~\cite{reda2020rainbows,reda2021color}. 
Kim et al.~\cite{Kim14} incorporate color aesthetics and contrast into the optimization of color assignment for image segments.
\yh{Szafir~\cite{szafir2017modeling} find that mark size heavily affects color discriminability.} 
Wang et al.~\cite{Wang2018} propose to maximize class discriminability based on color-based class separability, which takes into account spatial relationships between classes and in addition the contrast with the background color.
Once the assignment is done, the color of each class can be further optimized for better serving additional purposes, such as reducing the power consumption of displays~\cite{chuang2009energy},
improving the accessibility of visualizations for visually impaired users~\cite{machado2009physiologically}, or better class discrimination~\cite{lee2013perceptually}.
Almost all these methods aim to generate effective static visualizations, whereas our goal is to generate interactive visualizations with varying subsets of 
interest and maximizing class discriminability as well as the similarity between the perceived colors of each class.

\vspace{1.5mm}
\noindent\textbf{Color Palette Generation}.
To create an appropriate categorical color palette, the commonly used approach is to select one from a library of carefully designed palettes provided by online tools such as ColorBrewer~\cite{harrower2003colorbrewer}.
\yh{Fang et al.~\cite{fang2016categorical} suggest maximizing the perceptual distances among a set of colors while meeting various user-defined constraints.
Likewise, Nardini et al.~\cite{nardini2021automatic} provide an automatic optimization algorithm for improving continuous colormaps in Euclidean color space and integrate them into a test suite~\cite{nardini2020testing}.}
Colorgorical~\cite{Gramazio17} further allows users to customize color palettes by generating them based on user-specified discriminability levels and preferences.
Recently, Palettailor~\cite{Lu21} takes a further step by automatically generating categorical palettes for different types of charts, such as scatterplots or line and bar charts.
Rather than generating one palette at a time, our work produces a pair of contrastive palettes with different contrast over the background while maintaining color consistency between corresponding color pairs. This drives the dynamic generation of palettes to interactively highlight points of interest in multi-class scatterplots.

\vspace{1.5mm}
\noindent\textbf{Color Design for Highlighting}.
To let important classes stand out, the commonly used practice is to assign them bright colors while using subdued colors for less-important classes. This can be achieved by using
accentuated color palettes, which consist of a set of subdued colors and a set of bright (stronger, darker, or more saturated) colors. However, only few such palettes are available. For example, ColorBrewer~\cite{harrower2003colorbrewer} provides only a very small set of such palettes.
Therefore, Wilke~\cite{wilke2019fundamentals} suggests creating  palettes by lightening some colors of an existing palette while darkening others. This method might be able to create desired palettes but it is often hard to maintain the discriminability between all classes in the given data. In contrast, our method can automatically create such palettes and assigns them to the input data, while maintaining the stability of palettes for varying the data points of interest.

\yh{\vspace{1.5mm}
\noindent\textbf{Color Consistency.}
Multi-view visualizations are commonly used for multivariate analysis. Although a few design guidelines~\cite{wang2000guidelines} have been proposed for constructing multi-view visualizations, few of them are related to color design. Qu et al.~\cite{qu2017keeping} recommend a set of color consistency constraints across views.
Among them is a high-level constraint that the same data field should always be encoded in the same way. In our work, however, the highlighted data subset varies during the exploration.
To ensure a relative consistency of the perceived color, we require the highlighted and de-highlighted colors of the same class to have the same hue and similar color names.}

\section{Background}
Given a multi-class scatterplot with $m$ classes and $n$ data items $\mathbf{X}= \{\mathbf{x}_1, \cdots, \mathbf{x}_{n}\}$, each $\mathbf{x}_t$ has a label $l(\mathbf{x}_t)$ and the $i$-th class has $n_i$ data points. The goal of Palettailor~\cite{Lu21} is to find a color mapping  $\tau: L \mapsto c$ that maximizes the discriminability of the given multi-class scatterplot while ensuring that all colors can be referenced by names.
\yh{Since each class is assigned a unique color, a palette $P$ with $m$ colors is formed.}
Palettailor finds $\tau$ by  maximizing the following objective:
\begin{align}
	\mathop{\arg\max}_{\tau}E(\tau)= \omega_0 E_{PD}+\omega_1 E_{ND} + \omega_2 E_{CD},
\label{eq:objectivefunction-b}
\end{align}
consisting of a point distinctness term $E_{PD}$, a name difference term $E_{ND}$, and a color discrimination term $E_{CD}$. Each weight $\omega_i$ is a value range from 0 to 1 and
each class $C_i$ is assigned a unique color $c_i$. Besides these terms, a hard constraint is imposed to require the color difference between any two colors to be larger than a just noticeable difference threshold.  $E_{CD}$ is defined as the minimal CIELAB color distance~\cite{sharma2005ciede2000} among every color pairs in $\tau$; we describe the other terms in the following.


\begin{figure}[!t]
\centering
\includegraphics[width=0.98\linewidth]{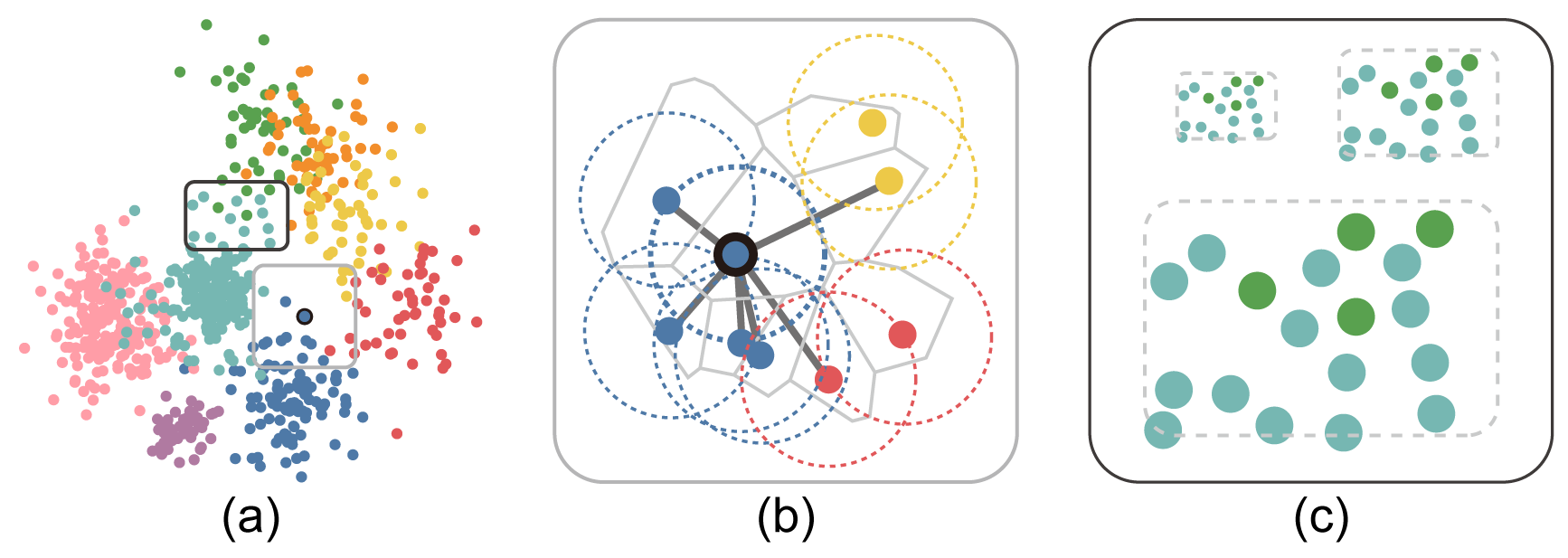}
\caption{\yh{Illustrating the nearest neighbors of each data point and the influence of mark size in the perceived color difference for an input scatterplot (a) with two  regions selected to zoom-in in (b,c). (b) The nearest neighbors for the selected point are defined by the $\alpha$-shape graph; (c) a subset of data points selected from (a) shown in three different sizes.}}
\vspace*{-3mm}
\label{fig:alphasize}
\end{figure}

\vspace{2mm}
\noindent\textbf{Point Distinctness}.
\yh{Given $\mathbf{X}$, an $\alpha$-shape graph~\cite{Lu21} is first constructed by connecting each point to its neighbors 
in the Delaunay graph and intersected within so-called $\alpha$-balls (see an example in Fig.~\ref{fig:alphasize}(b)).}
Then for each data point $\mathbf{x}_t$, its point distinctness is defined as:
\begin{align}
 \gamma (\mathbf{x}_t)=\frac{1}{|\Omega_t|} \sum_{\mathbf{x}_p \in \Omega_t}  \frac{\Delta\epsilon(\tau(l(\mathbf{x}_t)),\tau(l(\mathbf{x}_p)))}{d(\mathbf{x}_t,\mathbf{x}_p)} \nonumber ,
\end{align}
where $\Omega_t$ is set of nearest neighbors of $\mathbf{x}_t$, $\tau(l(\mathbf{x}_p))$ is the mapped color of $\mathbf{x}_p$,  $d$ is the Euclidean distance and $\Delta\epsilon$ is the CIELAB color distance~\cite{sharma2005ciede2000}. By summing up the point distinctness of all data points,   $E_{PD}$ is defined as:
\begin{align}\label{eq:pd}
 E_{PD} = \sum_i^m  \frac{1}{n_i}\sum^{n}_{t=1}\gamma(\mathbf{x}_t) \delta(l(\mathbf{x}_t),i),
\end{align}
where $\delta(l(\mathbf{x}_t),i)$ is one, if the class label $l(\mathbf{x}_t)$ has a value of $i$, otherwise zero. If a class overlaps with different classes, the point distinctness  value will be high, but small for a well-separated class.

\vspace{2mm}
\noindent\textbf{Name Difference.}\
Since color names are frequently used for communicating colors in visualization, a good palette should consist of colors associated with largely different names. Given a normalized color-term count matrix $T$, $E_{ND}$ is defined as:
\begin{align}\label{eq:nd}
 E_{ND} = \frac{2}{m(m-1)}\sum_{i \neq j \in m} D(T_{c_i},T_{c_j}).
\end{align}
where $T_{c_i}$ is the probability distribution of color names for a given color $c_i$ and $D(T_{c_i},T_{c_j})$ can be any distance measure for probability distributions; here we use the cosine distance~\cite{heer2012color}.

To find the optimal $\tau$ in Eq.~\ref{eq:objectivefunction-b}, a customized simulated  annealing~\cite{aarts1989stochastic} algorithm is used, which starts with a random initial solution and a high temperature and progressively updates the solution and decreases the temperature to zero until reaching the convergence.
This algorithm yields reasonable results \yh{in less than 10s for 20 classes}, facilitating an interactive generation of palettes. \yh{However, a key limitation of Palettailor (and other automated colorization techniques) is that they do not support interactive highlighting, limiting their use in dynamic visualizations.  We address this limitation by providing explicit support for the interactive emphasis on demand. Additionally, we also extend earlier colorization methods by modeling background contrast. This allows the highlighted data to pop out  relative to the background, but also to other marks of the visualization. We also developed new constraints for the optimization to ensure discriminability and color consistency for emphasized and non-highlighted classes.} 

\section{Context-Preserving Highlighting}
For a given multi-class scatterplot with $m$ classes and $n$ data items $\mathbf{X}$ and a background color $c_b$, our goal is to find a set of colors 
that creates the desired interactive emphasis effect for multi-class scatterplots. In line with the design requirements for pop-out effects and categorical data visualization~\cite{Itti98, Gleicher18, Lu21},
our problem can be formulated based on the following three design requirements:
\begin{enumerate}
\item[(i)] \textbf{DR1:} highlighting the \yh{\od{seletced}
 data points} as much as possible to deliberately attract user attention;
\item[(ii)] \textbf{DR2:} maximizing the visual discrimination between classes for efficiently exploring the data, for the selected and non-selected classes; and
\item[(iii)] \textbf{DR3:} maintaining \yh{color consistency for data points  \od{when they are dynamically highlighted or de-emphasized}.}
\end{enumerate}
The resulting color mapping schemes  satisfy DR1 by letting selections of interest pop out from the context while yielding  better visual discrimination of classes for meeting DR2. \yh{Because the color for a class can vary depending on whether it is highlighted or not, we satisfy DR3 by requiring the two states \od{(highlighted vs. de-emphasized)} to have the same hue and saturation values while also ensuring similar color names. In doing so, we ensure a consistent perception of color appearance as data points of interest are interactively highlighted or de-emphasized.}

\subsection{Combination-based Highlighting}
Most existing colorization techniques~\cite{Gramazio17, Lu21} attempt to meet DR2. A key challenge for our technique, however, is to ensure that colors for classes of interest are sufficiently distinct to create \od{the wanted} pre-attentive `pop out' effect (DR1).
To meet this constraint, a widely used manual approach~\cite{munzner2014visualization} is to modulate the color opacity or luminance contrast with the background for the non-selected data points.
However, doing so will likely violate DR2 and DR3, because changing one or multiple colors might not preserve the ability of the viewer to visually discriminate all classes (see the bottom row of Figs.~\ref{fig:teaser} (b,c)).
\yh{On the other hand, simply extending existing colorization methods to enforce larger color differences for all classes might help meet DR1 and DR2, but the generated color mappings might differ noticeably when data points are dynamically emphasized (e.g., in response to user selection), causing user confusion (violating DR3).} 

\yh{To meet the three design requirements, we propose a combination-based highlighting method consisting of two steps.
We first pre-generate the two color mapping schemes $\tau^s: L \mapsto c$ and $\tau^f: L \mapsto \tilde{c}$ for a given scatterplot, consisting of one palette $P^s=\{c_1, \cdots, c_m\}$ with salient colors over the background and a corresponding palette $P^f=\{\tilde{c}_1, \cdots, \tilde{c}_m\}$ with faint colors.  For each data point $\mathbf{x}_i$,  the color will be $\tau^s(l(\mathbf{x}_t))$ if selected for highlighting or $\tau^f(l(\mathbf{x}_t))$ otherwise.
Since $\tau^s$ and $\tau^f$ both are required to meet DR2 and DR3, the overall color mapping will emphasize the data of interest (given the high saliency of $\tau^s$) while also preserving the visual discriminability of all classes and ensuring good color consistency.}

\subsection{Modeling Contrastive Color Mappings}
We formulate the search for \revised{a pair of color mapping $\tau^s$ and $\tau^f$ and their resulting palettes $P^s$ and $P^f$ as an optimization problem with the \textbf{objective function} $E(\tau^s, P^s, \tau^f, P^f)$}:
\begin{align} 
	\mathop{\arg\max}_{\tau^s, P^s, \tau^f, P^f}E(\tau^s, P^s, \tau^f, P^f)= &\omega_0 \big(E_{PD}(\mathbf{X},\tau^s)+ E_{PD}(\mathbf{X},\tau^f)\big) \\
+&\omega_1  \big(E_{BC}(\mathbf{X},\tau^s) -   E_{BC}(\mathbf{X},\tau^f) \big)\\
+&\omega_2 \big(E_{ND}(P^s) + E_{ND}(P^f)\big) \\
+&\omega_3 \ E_{CC}(P^s,P^f)
\label{eq:objectivefunction}\vspace{-2mm}
\end{align}
where the first two terms are based on the score of each color mapping scheme and the last two measure the score and compatibility of two resulting color palettes.

Each weight $\omega_i$ is a value in the range $[0,1]$; we set all of them to 1 by default. The terms $E_{PD}$ and $E_{ND}$ are designed to meet DR2 while ensuring that colors are nameable.
The second term $E_{BC}$ satisfies DR1 by maximizing and minimizing the \yh{luminance} contrast of the \revised{color mappings $\tau^s$ and $\tau^f$} over the background, respectively.
The last term $E_{CC}$ meets DR3 by requiring the corresponding colors in the two palettes to have similar perceived colors.
\yh{To ensure all colors found by color mapping $\tau$ have enough discriminability, we apply a hard constraint in the form of the JND threshold.
Since the perceived color difference varies across mark sizes (see Fig.~\ref{fig:alphasize}(c)), we define the JND threshold based on the size-dependent model proposed by Szafir\cite{szafir2017modeling}.} $E_{PD}$ and  $E_{ND}$ are defined in Eq.~\ref{eq:pd} and Eq.~\ref{eq:nd}.
In the following, we will introduce the new terms for background contrast $E_{BC}$ and color \yh{consistency} $E_{CC}$ and then describe how we solve the overall optimization problem.

\vspace{2mm}
\noindent\textbf{Background Contrast}.
\yh{We define the contrast of each data point $\mathbf{x} _t$ to the background based on two factors:  position-based class separability among its neighboring points and luminance difference to the background.}
The former measures by the difference between two separation degrees~\cite{Aupetit02}:
\begin{align}
\rho(\mathbf{x} _t)&= b(\mathbf{x} _t)-a(\mathbf{x} _t) \ ,\nonumber
\end{align}
where $b(\mathbf{x}_t)$ is the between-class separation degree and $a(\mathbf{x}_t)$ is the within-class separation degree.
The measures are defined as weighted sums of the non-separability of $\mathbf{x}_t$ from its neighborhood stemming from the same class and from other classes:
\begin{align}
a(\mathbf{x}_t)&=\frac{1}{|\Omega_t|}\sum_{\mathbf{x}_p \in \Omega_t } \frac{\delta(l(\mathbf{x}_t), l(\mathbf{x}_p))
}{d(\mathbf{x}_t,\mathbf{x}_p)} ,\nonumber 
\end{align}
\begin{align}
b(\mathbf{x}_t)=\frac{1}{|\Omega_t|}\sum_{\mathbf{x}_p \in \Omega_t } \frac{1-\delta(l(\mathbf{x}_t), l(\mathbf{x}_p))}{d(\mathbf{x}_t,\mathbf{x}_p)}  .\nonumber
\end{align}

When most neighbor points of $\mathbf{x}_t$ have the same label as $\mathbf{x}_t$,  $\rho (\mathbf{x}_t)$ is negative.
However, a negative $\rho (\mathbf{x}_t)$ \yh{would reduce the contrast of the palette $P^s$  over the background}, which
conflicts with the objective of Eq.~\ref{eq:objectivefunction}.
To address this issue, we 
use an exponential function to let $\rho (\mathbf{x}_t)$  always be positive and then normalize it to the range [0,1]. 
The contrast to the background of the $i$th class is:
\begin{align}\label{eq:ctbc}
 \beta_i(\mathbf{X},\tau) = \frac{1}{n_i}\sum^{n}_{t=1} \rho(\mathbf{x}_t)\yh{\Delta L(\tau(l(\mathbf{x}_t)),\mathbf{c}_b)} \delta(l(\mathbf{x}_t),i).
\end{align}
where $\tau(l(\mathbf{x}_t)$ is the color of point $\mathbf{x}_t$, \yh{$\Delta L$ is the absolute luminance difference between point and background color in CIELAB space}. The background contrast is defined as the sum of the background contrasts of each class:
\begin{align}\label{eq:Epd}
 E_{BC}(\mathbf{X},\tau) = \sum_{i=1}^{m} \beta_i (\mathbf{X},\tau).
\end{align}

\begin{figure}[!ht]
\centering
\includegraphics[width=1\linewidth]{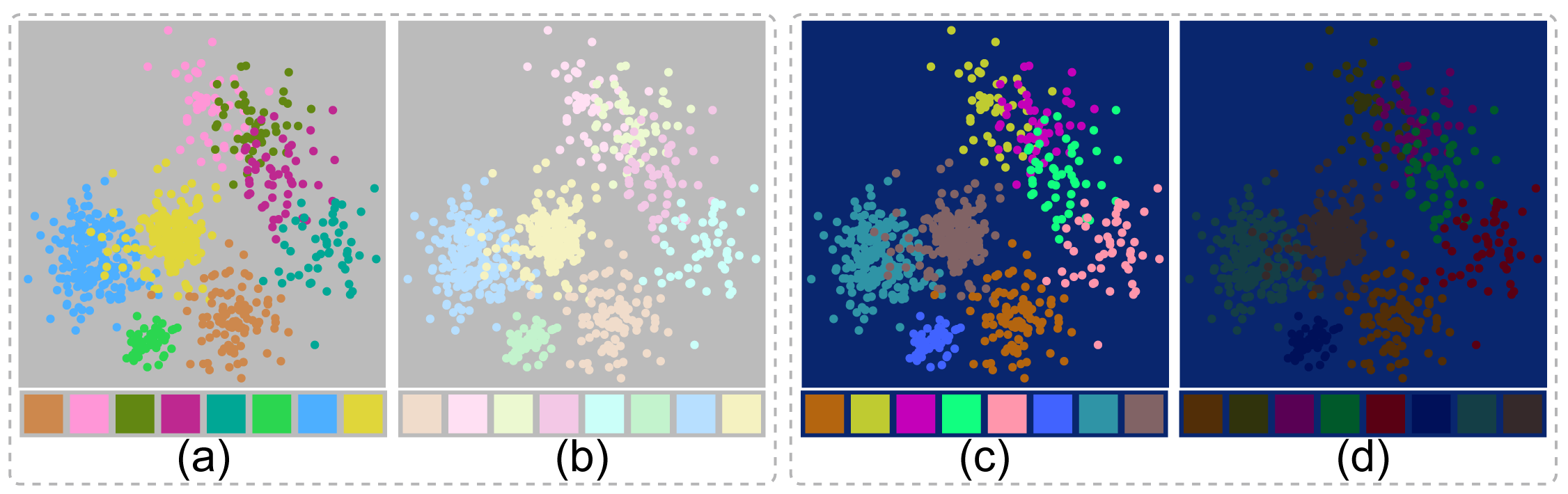}
\caption{Results generated for two different backgrounds: grey (a,b) and blue (c,d). (a,c) the salient palettes and the colorized scatterplots; (b,d) the faint palettes and the colorized scatterplots. 
}
\label{fig:background}
\end{figure}

%

This term depends on the non-separability and \yh{color luminance} difference to the background, which means a class overlapping with other classes should have a larger $\beta_i$ than a separated class. As shown in the top of Fig.~\ref{fig:teaser}(d) (green and dark blue classes), this yields better class separability.
Since we take the contrast with the input background color into account, this model is able to adapt to different backgrounds, see Fig.~\ref{fig:background} for illustration.

\vspace{2mm}
\noindent\textbf{Color \yh{Consistency}}.
To ensure colors are perceived similarly for highlighted and de-emphasized states, hue and saturation should be stable when data emphasis is dynamically changed. We also require colors to have similar color names, which helps the user in maintaining a mental map as data is selected and deselected. Hence, the following term measures name similarity between color pairs across the two contrastive palettes:
\begin{align}
E_{CC}(P^s,P^f)  &= -\frac{1}{m}\sum_{i \in m}ND(P^s(i),P^f(i)), \\
 \text{subject to} \quad
  &H(P^s(i)) = H(P^f(i)) \mbox{ and }  S(P^s(i)) = S(P^f(i)) ,\label{eq:hard} \\
   &\mbox{for } i = 1, \dots, m \nonumber 
\end{align}

\begin{figure}[!t]
\centering
\includegraphics[width=1\linewidth]{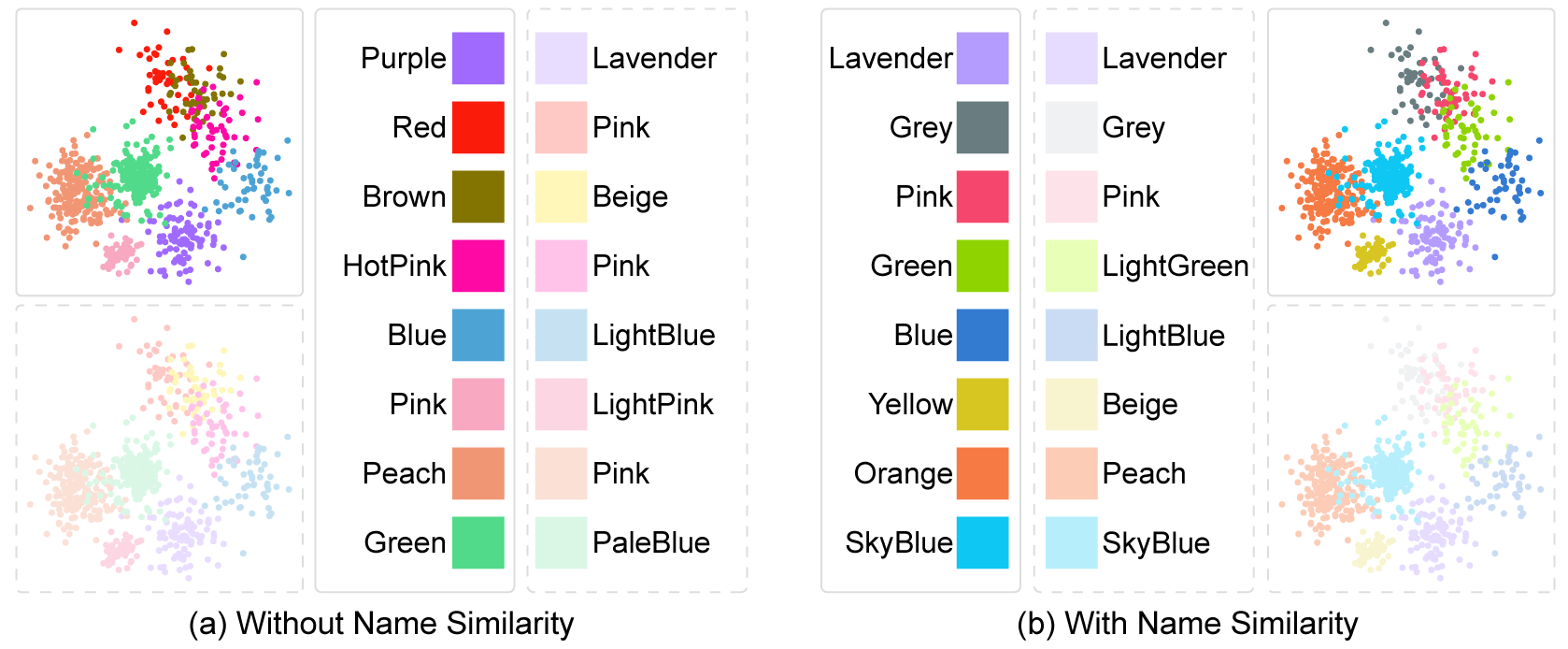}
\caption{Comparing palettes generated by without (a) and with (b) name similarity terms. The results with solid borders are created by the salient palettes and the ones with the dashed border are created by the faint palettes.}
\vspace{-3mm}
\label{fig:nameConsistency}
\end{figure}

We impose the hard constraint that $P^s(i)$ and $P^f(i)$ assigned to the $i$th class should have the same hue $H(P^s(i))$ and saturation $S(P^s(i))$ values. In other words, the difference between colors in $P^s(i)$ and $P^f(i)$ is only in the lightness channel and thus the objective in Eq.~\ref{eq:objectivefunction} can be simplified to a four-dimensional optimization problem.
Fig.~\ref{fig:nameConsistency} compares the results generated by only imposing the hard constraint in Eq.~\ref{eq:hard} and the complete color consistency term.
Incorporating name similarity not only yields contrastive palettes with highly similar color names but also enlarges  name differences in each palette (see the four pink colors of the right palette in Fig.~\ref{fig:nameConsistency}(a)).

\yh{
\vspace{2mm}
\noindent\textbf{Foreground Contrast Constraint}.
To ensure a visual pop-out effect for the highlighted data, we require the corresponding points to have large color contrast to the background, so that highlighted data is perceived as a clear `foreground' layer. Since highlighted points are colorized by the $P^s$ palette, we meet this requirement by imposing a hard constraint that
each color in $P^s$ should have a larger luminance contrast to the background than all colors in $P^f$:
\begin{align}\label{eq:lc}
\forall i, \forall j, \Delta  L(P^s(i),\mathbf{c}_b) > \Delta L(P^f(i),\mathbf{c}_b).
\end{align}
For short, we refer to this constraint as the foreground contrast constraint.}

\vspace{2mm}
\noindent\textbf{Homogeneous Lightness Constraint}.
Previous studies~\cite{ware2019information} show that detecting targets over uniform backgrounds is more efficient than over complex ones. Thus, assigning uniform lightness to all colors in $\tau^f$ allows us to better meet DR1. However, this might reduce visual discrimination among non-highlighted classes, which would violate DR2. \yh{Fig.~\ref{fig:sigma}(b) shows an example with uniform lightness; note how the assigned pink and red colors in Fig.~\ref{fig:sigma}(a) become very similar in Fig.~\ref{fig:sigma}(b).}
To find a trade-off between these two requirements, we impose a constraint that all colors in $P^f$ should have a small standard deviation in lightness so as to yield a relatively homogeneous background:
\begin{align} \label{eq:hb}
0 \leq \text{SD}(\{L(P^f(1)), \dots, L(P^f(m))\}) \leq \sigma 
\end{align}
where $L(P^f(i))$ is the lightness of color $P^f(i)$ and $\sigma$ is a small value specified by the user. \yh{To meet this constraint, we first find a uniform lightness level for all colors in each $\tau$ and subsequently perturb the lightness of each color $P^f(i)$  within a range of $[-\sigma, \sigma]$. Figs.~\ref{fig:sigma}(c, d) show how an appropriate $\sigma$ can help to meet DR1 and DR2. Note how all colors in Fig.~\ref{fig:teaser}(d) are discriminable (see \S\ref{subsec:parameter} for a formal analysis of this parameter). Also, observe how the non-highlighted points are assigned similarly faint colors though not identical lightness levels.}




\begin{algorithm}[!thb]
\caption{Simulated Annealing for Palette Generation}
\begin{algorithmic}[1]
\State randomly initialize $P^s$ and $l$
\State set an initial temperature $T$ and $P^f$ = $P^s$
\State set the lightness of all colors in $P^f$  to $l$
\State $\Delta E = 0$
\While{$T>0$}
\yh{
\State $Q^s = P^s, Q^f = P^f$
\If{$random(0,1) < exp(\Delta E /T)$}
\State randomly choose a new $\hat{l}$ in the neighborhood of $l$
\State set the lightness of all colors in $Q^f$ to $\hat{l}$
\EndIf
\If{$random(0,1) < 0.5$}
\State randomly exchange two colors from $Q^s$
\State exchange the corresponding colors in $Q^f$
\Else
\State randomly disturb one color $c$ from $Q^s$
\State update $Q^f$ with $Q^s$ via Eq.~\ref{eq:hard} and Eq.~\ref{eq:hb}
\While{$ Q^s, Q^f$ not satisfying $Eq.~\ref{eq:lc} $}
\State  disturb color $c$ from $Q^s$
\State  update $Q^f$ with $Q^s$ via Eq.~\ref{eq:hard} \yh{and Eq.~\ref{eq:hb}}
\EndWhile
\EndIf
}
\While{$\displaystyle \mathop{\min}_{c_i, c_j \in Q^s }\Delta \varepsilon(c_i, c_j)< \eta $ \mbox{or}
 $\displaystyle \mathop{\min}_{c'_i, c'_j \in Q^f }\Delta \varepsilon(c'_i, c'_j)< \eta $
}
\State  randomly disturb $c_i$ or $c_j$ to get a new $Q^s$
\State  update $Q^f$ with $Q^s$ via Eq.~\ref{eq:hard} \yh{and Eq.~\ref{eq:hb}}
\EndWhile
\State $\Delta E=E(Q^s, Q^f)-E(P^s, P^f)$
\If{$\Delta E>0$}
\State $P^s = Q^s$, $P^f = Q^f$, $l=\hat{l}$
\Else
\If{$random(0,1) < \exp(\Delta E /T)$}
\State $P^s = Q^s$, $P^f = Q^f$, $l=\hat{l}$
\EndIf
\EndIf
\State $T = \alpha T$
\EndWhile

\end{algorithmic}
\label{alg:sa}
\end{algorithm}

\vspace{-2mm}
\subsection{Optimization for Contrastive Palettes }
\label{sec:optimization}
We implement the above constraints in a simulated annealing algorithm to generate
\revised{a pair of color mapping schemes $\tau^s$ and $\tau^f$ (see pseudocode in Algorithm~\ref{alg:sa}).
Before presenting our algorithm, we map each class label to an index range of $[1,m]$ and assume that the color $c_i$ in the palette $P$ is assigned to the $i$th class.}
After initializing a high ``temperature'' and a  palette $P^s$ with $m$ random colors, this method iteratively updates the palettes with three major steps in each iteration: i) finding a neighboring solution of $P^s$, ii) using $P^s$ to update $P^f$ and finding a neighboring solution of $P^f$; and iii) refining the two temporary palettes $Q^s$ and $Q^f$ to meet the JND constraint.
In the following, we describe the last two major steps.

\begin{figure}[!t]
	\centering
	\includegraphics[width=1\linewidth]{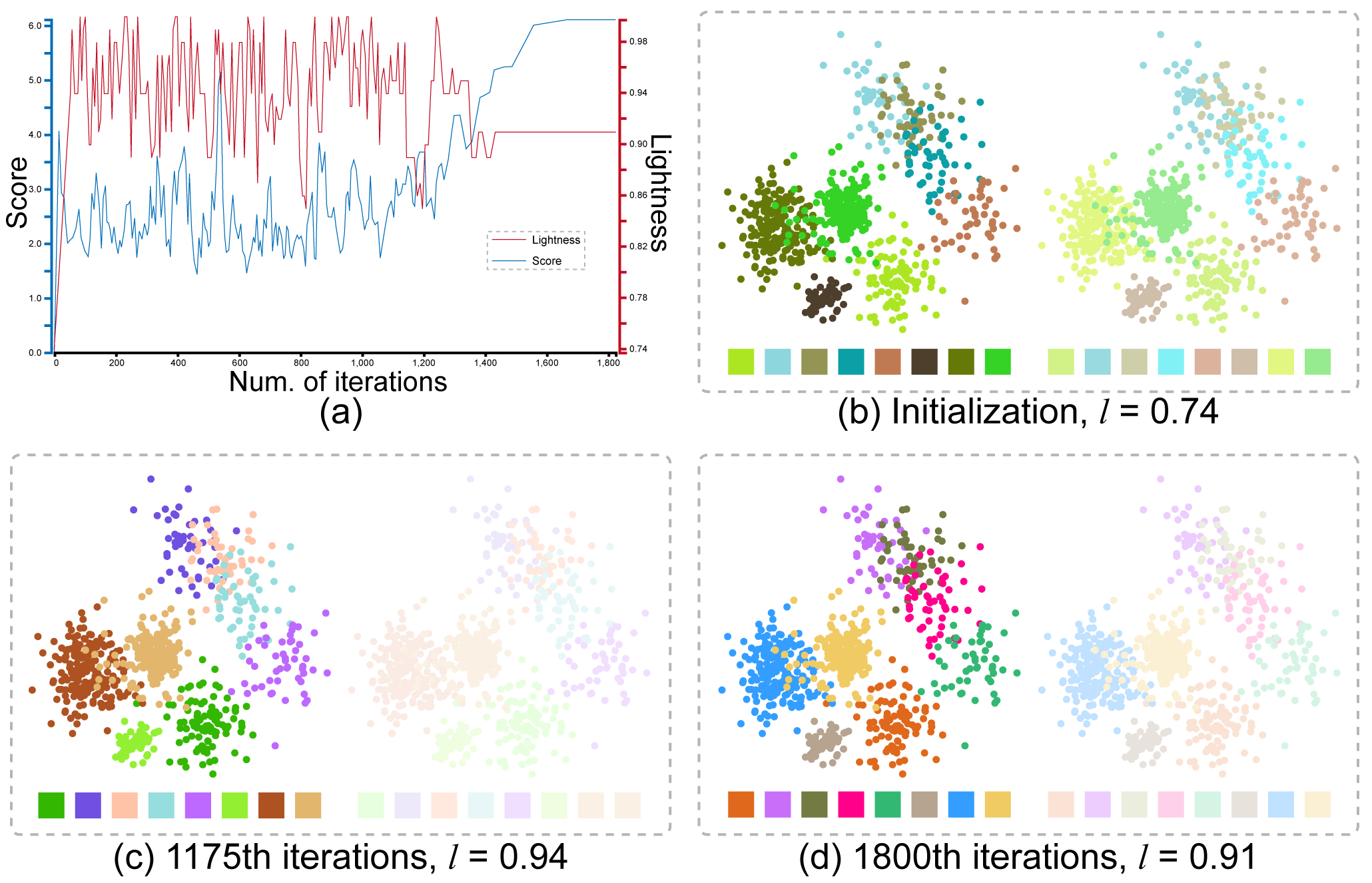}
	\caption{Convergence of our method. (a) curve on $E$ versus the number of iterations(blue) and
  curve of $l$ versus the number of iterations (red); (b) colorized results and palettes with random initialization; (c,d) results after 1175 and 1800 iterations.
	}
	\label{fig:timeVSclassnum}
\vspace{-3mm}
\end{figure}

\vspace{2mm}
\noindent\textbf{Finding the neighboring $Q^f$ near $P^s$ (line 6-21).}
\yh{Based on the current solution $P^s$ and $P^f$, we first update $Q^f$ by finding a new lightness value for all colors (line 6-10). To rapidly produce a homogeneous background, we set a large probability for accepting a uniform lightness for all colors initially and decrease it with increasing number of iterations (see the probability distribution in the supplemental material).
Then, we try to find a neighbor solution $Q^s$ of $P^s$ by randomly exchanging two colors (line 12) or choosing a new color around the neighborhood of one selected color (line 15). After generating a new solution for $Q^s$, we update $Q^f$ in the terms of Eq.~\ref{eq:hard} and Eq.~\ref{eq:hb} (line 16). Namely, we maintain the hue and saturation of each color chosen from $Q^s$ and perturb the latest uniform lightness value (see lines 28 and 31) by a random value in  $[l-\sigma, l+\sigma]$ to increase the discriminability. Finally, we disturb the solution until satisfying Eq.~\ref{eq:lc} (see line 17-20).
}

Fig.~\ref{fig:timeVSclassnum} (a) shows the evolution curve of $l$, which has large variations at the beginning and then gradually converges to a stable value. Figs.~\ref{fig:timeVSclassnum}(b, c, d) show scatterplots visualized with palettes yielded at the initialization, the 1175th, and the 1800th iteration.
$l$ quickly reaches a reasonable value but the colors keep changing to further improve class discrimination (see Figs.~\ref{fig:timeVSclassnum}(c, d)).

\vspace{2mm}
\noindent\textbf{Refine Palettes via JND constraints (line 22-25).}\
To ensure that all colors in $Q^s$ and $Q^f$ can be discriminated, we compute the minimal distance between each pair of colors in both palettes and see if it satisfies the hard JND constraint. If their difference is smaller than a \yh{mark size-dependent} JND threshold $\eta$~\cite{szafir2017modeling}, we randomly perturb the corresponding color pair in $Q^s$ and use $Q^s$ to update $Q^f$ until all color pairs meet the constraint. \rereview{Since $\eta$ is a JND threshold modulated by the given mark size, we recalculated $\eta$ when the size is changed. JND modulation ensures that smaller marks are allocated larger color difference relative to other classes for discriminability.}

Due to the stochastic nature of this algorithm, random initialization of palettes and lightness does not influence the final solution in our experiments.
The time complexity of each iteration is $O(m^2)$ and the time complexity for the whole algorithm is $O(tm^2)$ for a total number of iterations $t$.

Our method allows us to yield reasonable palettes for scatterplots with 20 classes in less than 10s.

\subsection{Parameter Analysis for Background Complexity}
\label{subsec:parameter}

\begin{figure}[!htb]
\centering
\includegraphics[width=1\linewidth]{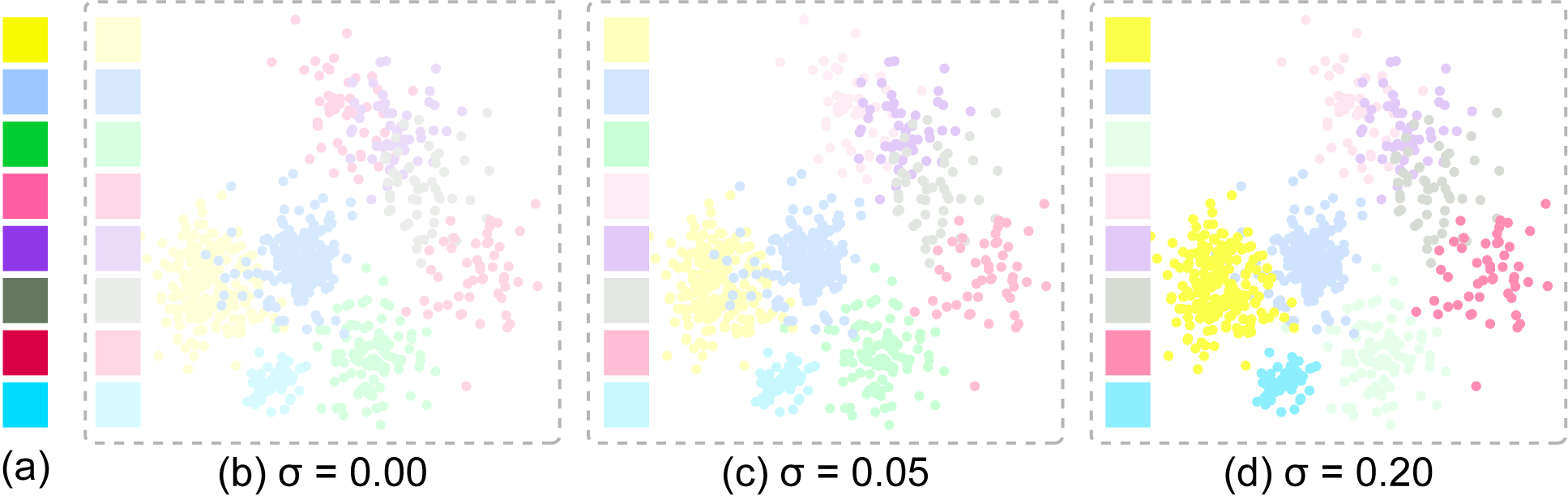}
\caption{Effect of the lightness standard deviation $\sigma$ on the faint palette: (a) salient palette $P^s$;
 (b,c,d) faint palettes $P^f$ generated by different $\sigma$ (left) and resulting scatterplots (right).}
\vspace*{-3mm}
\label{fig:sigma}
\end{figure}

A key parameter that affects the likelihood that the emphasized data will pop out is the complexity of the background. Namely $\sigma$, which controls the standard deviation for the lightness of the non-highlighted data (i.e., for the colors of the faint palette $P^f$). A small $\sigma$ will reduce the discriminability of all background classes (see the two red colors in Fig.~\ref{fig:sigma}(b)), while a large value will degrade user performance in identifying the highlighted data (see the yellow and pink classes in Fig.~\ref{fig:sigma}(d)). Experimenting with different levels, we found a default value of $\sigma = .05$ to be a good trade-off between emphasizing classes of interest while preserving background discriminability (see Fig.~\ref{fig:sigma}(c)).


\section {Evaluation}
\label{sec:evaluation}

Considering that oftentimes there are multiple tasks involved in the analysis of a scatterplot, we evaluated the effectiveness of our method from two different perspectives: \emph{static visualization} and \emph{interactive exploration}. For static visualizations, all classes have equal importance and the main task is to discriminate different classes (see Table~\ref{tab:summaryTask}). Here, we use a \revised{\emph{counting task}, prompting participants to count the number of unique classes in a visualization, thus measuring how discriminable the classes are}. For interactive exploration, classes of interest should be highlighted and other classes need to be de-emphasized. In this case, the most important part is to find the classes of interest. However, during  interactive exploration, different classes would be highlighted. When a class gets out of focus and its color faints, the viewer still needs to recognize which class it is. Furthermore, people often need to distinguish the context around a highlighted class to examine the data distribution. To achieve these evaluation goals, we created three tasks to examine the efficiency of our method: (1) a \emph{highlighting task}, (2) a \revised{\emph{matching task}} and (3) a \revised{\emph{selecting task}. The last two  are designed to evaluate the effectiveness of our technique in preserving color consistency and class discrimination of non-selected data points, referred as context-preserving tasks.}
We conducted two controlled experiments across the four tasks by crowdsourcing 
150 participants through Amazon Mechanical Turk (AMT).

\begin{table*}[htbp]
\centering

\caption{
\revised{
 Layout of the two crowdsourced experiments: \emph{Tableau (D)} indicates \emph{Tableau with default assignment}, \emph{Tableau (O)} indicates \emph{Tableau with optimal assignment}, \emph{Palettailor (L)} indicates \emph{Palettailor with lightness adjustment}, \emph{Palettailor (A)} indicates \emph{Palettailor with alpha blending}, \emph{Tableau (D+H)} indicates \emph{Tableau Highlighter with default assignment}, \emph{Tableau (O+H)} indicates \emph{Tableau Highlighter with optimal assignment}, \emph{Our Method (S)} indicates \emph{Our Method (static)}, \emph{Our Method (I)} indicates \emph{Our Method (interactive)}.
}
}
\resizebox{0.98\linewidth}{!}{
\begin{tabular}{c|c||c|c|c|c}
 \hline
 \multicolumn{2}{c||}{\textit{{Experiment 1: Static Visualization}}} & \multicolumn{4}{c}{\textit{{Experiment 2: Interactive Exploration}}}\\
 \hline
 \textit{Methods} & \textit{Tasks} & \textit{Methods} & \multicolumn{3}{c}{\textit{Tasks}}\\
 \hline
 \multirow{2}{*}{ Palettailor} &  & Palettailor (L) & & &  \\

   &  & \revised{Palettailor (A)} & & &  \\

  \revised{Tableau (D)} & \makecell[c]{\revised{Counting Task}} & \revised{Tableau (D+H)} & \makecell[c]{Highlighting Task} & \makecell[c]{\revised{Matching Task}} & \makecell[c]{\revised{Selecting Task}} \\

 Tableau (O) &  & Tableau (O+H) & &  &  \\

 \makecell[c]{Our Method (S)} &  & \makecell[c]{Our Method (I)} & & &  \\
 \hline
\end{tabular}
}

\label{tab:summaryTask}
\vspace{-3mm}
\end{table*}

\vspace{2mm}
\noindent{\textbf{Benchmark Methods.}}
We compared our method with two existing colorization methods as benchmarks: 1) Palettailor~\cite{Lu21}, a state-of-the-art automated colorization tool designed for generating discriminable and optimized categorical palettes; and 2) Tableau~\cite{tableau}, an interactive data visualization software with  designer-crafted palettes. 
Tableau also has a purposefully designed tool called Highlighter for emphasizing a specific class while maintaining the context of the other classes. We therefore include Tableau Highlighter among our benchmark methods. However, Palettailor only supports static data visualization. Hence, we compared our method with two commonly used extrinsic emphasis techniques: adjusting the lightness contrast or \revised{the opacity value of a given palette to emphasize desired classes}~\cite{liang2010highlighting}.

\begin{figure}[hbt]
	\centering
	\includegraphics[width=1\linewidth]{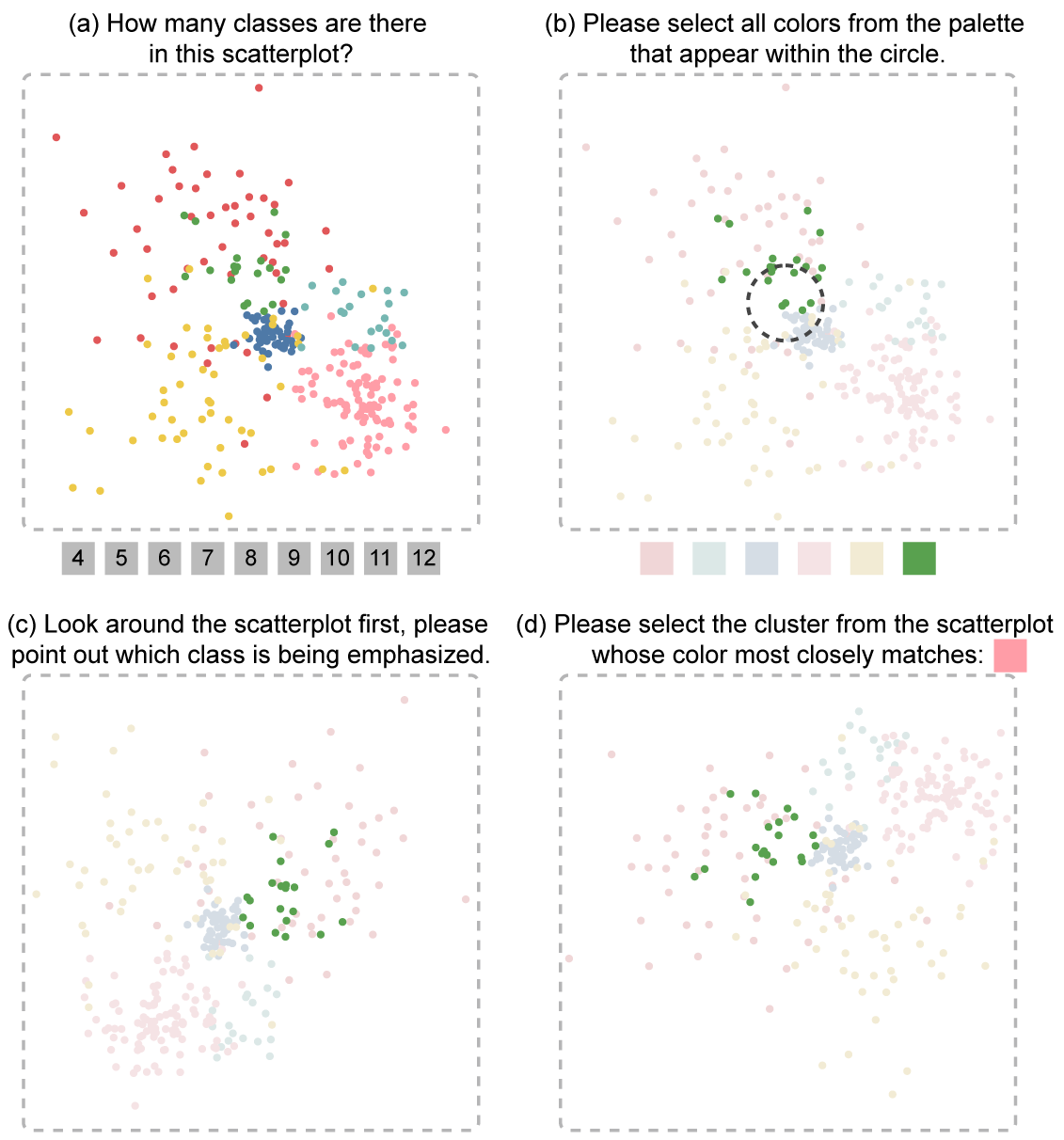}
	\caption{Illustration of tasks: (a) \revised{\emph{counting task}}; (b) \revised{\emph{selecting task}}; (c) \emph{highlighting task} and (d) \revised{\emph{matching task}}. \revised{The scatterplot is from one of the datasets used in the experiment.  Here, we apply a Tableau-provided color palette} with randomized rotation to avoid learning effects.}
	\vspace*{-3mm}
	\label{fig:tasks}
\end{figure}

\vspace{2mm}
\noindent{\textbf{Tasks \& Measures.}}
\begin{itemize}
\setlength{\itemsep}{5pt}
    \item \revised{\emph{Counting task (global discrimination).}}
        Following previous methodologies~\cite{Wang2018, Lu21}, we asked participants to identify how many classes (i.e., different colors) they can see in a given scatterplot. As shown in Fig.\ref{fig:tasks}(a), they entered their answer by selecting from multiple options that were displayed below the scatterplot. We recorded the answer and \emph{response time} for each trial, and computed the relative \emph{error} as the proportion of the total number of classes. For example, a participant answering with 8 classes when  there were actually 10 would be reported as an error of 0.2. 
    \item  \revised{\emph{Selecting task (local discrimination).}}
        As shown in Fig.\ref{fig:tasks}(b), we put a circle around a randomly selected point of the highlighted class in a scatterplot and then asked participants to select all colors from the palette that appear within the circle. The radius of the circle was $10 \times $ the radius of each point. Participants entered their answers by selecting colors from a palette displayed below the scatterplot. We recorded user selection and \emph{response time} for each trial, and computed the relative \emph{error} as the proportion of the actual number of classes. 
    \item \emph{Highlighting task.}
        Following the methodology of Mairena et al.~\cite{mairena2021emphasis}, we asked participants to examine the scatterplot first and then point out which class they believe is being emphasized, as shown in Fig.~\ref{fig:tasks}(c). We allow participants to click any representative point in the scatterplot as a way of selecting the class.
        For each trial, we measured the \emph{binary error} (i.e., whether the class chosen by a participant is the intentionally highlighted one). We also tracked the \emph{response time}.
    \item  \revised{\emph{Matching task (color consistency).}}
        As shown in Fig.\ref{fig:tasks}(d), we asked participants to select the cluster from a scatterplot whose color most closely matches the indicated color. Participants could click any representative point in the scatterplot as a way of selecting the class.
        For each trial, we measured the \emph{error (0/1)} (i.e., whether the class chosen by a participant was the correct one) and the \emph{response time}.
\end{itemize}


\vspace{.3em}
\noindent{\textbf{Hypotheses.}}

\revised{We expect our methods to outperform the benchmarks in preserving context and color consistency. That is, we expect to attain the benefits of visual focus without sacrificing performance on tasks requiring context. }Specifically, we pose the following hypotheses:

\begin{itemize}[noitemsep]
\setlength{\itemsep}{5pt}
    \item[\textbf{H1.}] \revised{Our palette generation method for static visualization is comparable to the benchmark conditions in the \emph{counting task}}.
    \item[\textbf{H2.}] \revised{Our palette generation method for interactive exploration is comparable to the best benchmark conditions in the \emph{highlighting task}}. 
    
    \item[\textbf{H3.}] \revised{Our palette generation method for interactive exploration outperforms the benchmark conditions in the two context-preserving tasks (\emph{selecting task} and \emph{matching task})}.
    
\end{itemize}

\vspace{.3em}
\noindent{\textbf{Dataset Generation.}}
The scatterplot datasets used in our studies were generated as follows.
First, to avoid learning effects, we chose three different class numbers: 6, 8 and 10 classes. Each class was generated using Gaussian random sampling and random placement in a $600 \times 600$ area.
Following the procedure described in Lu et al.~\cite{Lu21}, all scatterplots belonged to one of four possible configurations of class size and density: small \& dense ($n=50, \sigma=20$), small \& sparse ($n=20, \sigma=50$),  large \& dense ($n=100, \sigma=50$), and large \& sparse ($n=50, \sigma=100$). In total, we generated 3 (class number) $\times$ 4 = 12 scatterplots.

\vspace{.3em}
\noindent{\textbf{Engagement Checks.}}
In addition to the analyzed trials, we also generated multiple engagement checks to verify that participants were paying attention to the task. Engagement checks comprised a scatterplot with only 4 fully-separated classes, each with a very distinctive color. For the \emph{highlighting task}, we randomly chose one class to be emphasized, and assigned the other classes a lightness value of 0.9 to let them fade into the white background. For the other tasks, we assigned highly distinctive colors to allow for an easy class discrimination. We excluded participants from the analysis  who failed more than one engagement check.

\vspace{.3em}
\noindent{\textbf{Procedure. }}
Each participant went through the following steps: (i) viewing an instruction for the task and completing three training trials; (ii) completing each analyzed trial as accurately as possible; (iii) providing demographic information.
The three training trials were identical to the subsequent real test.
We implemented different response mechanisms for the four tasks. For the highlighting and matching tasks, participants clicked a data point belonging to the class they thought was the correct one. For the counting and selecting tasks, participants entered their class count or selected the corresponding colors by choosing from multiple options displayed below the visualization.

\vspace{.3em}
\noindent{\textbf{Analysis. }}
Following previous research~\cite{Lu21}, we analyzed the results using 95\% confidence intervals, and conducted Mann-Whitney tests to compare the differences between the conditions. In addition, we computed the effect size using \emph{Cohen's d} (i.e., the difference in means of the conditions divided by the pooled standard deviation). We calculated an ANOVA-type statistics (ATS) without normality assumption (using the R-package GFD~\cite{gfd2017}) to examine the interaction effect between variables.


\subsection{Experiment 1: Static Visualization}
\label{subsec:Exp1}
We conducted this experiment to examine how well our method supports people to visually distinguish different classes in a static visualization through a \revised{\emph{counting task}}.

\vspace{.3em}
\noindent{\textbf{Conditions.}}
In this experiment, we included four conditions:
\begin{enumerate}
     \item \emph{Palettailor:}
        This method represents the state-of-the-art automated colorization algorithm for multi-class scatterplots with the best class discriminability, corresponding to Fig.~\ref{fig:teaser}(b)-top.

     \revised{\item \emph{Tableau with default assignment:}
        This method represents the default visualization effect for designer-crafted categorical palettes. We assigned each color to each class in turn, to mimic how Tableau performs the color assignment, as shown in Fig.~\ref{fig:teaser}(a)-top.
}
     \item \emph{Tableau with optimal assignment:}
        This method represents the state-of-the-art for designer-crafted categorical palettes.
        We applied the optimal discrimination assignment approach~\cite{Wang2018} to the Tableau-10 palette, to mimic the best discriminable result from a manual selection of the user, as shown in Fig.~\ref{fig:teaser}(c)-top.

     \item \emph{Our method (static):}
        Assigning colors using a salient palette generated by our automated colorization method described in Sec.~\ref{sec:optimization} with default settings. This condition reflects the fully-automated colorization option of our method for static visualization, as shown in Fig.~\ref{fig:teaser}(d)-top.
\end{enumerate}

\vspace{.3em}
\noindent{\textbf{Experimental Design.}}
We used a \emph{within-subject} design: each participant completed all four conditions.
To avoid ordering effects, we randomly shuffled the display order of the given 48 stimuli (4 conditions $\times$ 12 scatterplots). For each stimulus, we also randomly rotated the scatterplot. Furthermore, we added three engagement checks to ensure participants were paying attention to the experiment.

\vspace{2mm}
\subsubsection{\revised{Counting task}}
~\\
We asked participants to identify how many classes (i.e., distinct colors) they find in a given scatterplot, as shown in Fig.\ref{fig:tasks}(a). Participants choose an answer from multiple options given below the scatterplot. We expected to see that our method will be comparable to other state-of-the-art conditions w.r.t. error and response time. We conducted this task through AMT with 30 participants.
According to the completion time in the study (the details can be found in the supplementary materials), we paid each participant \revised{\$1.75} for the task based on the US minimum hourly wage.
No participant claimed color vision deficiency on their informed consent.

\begin{figure}[htb]
\centering
\includegraphics[width=1\linewidth]{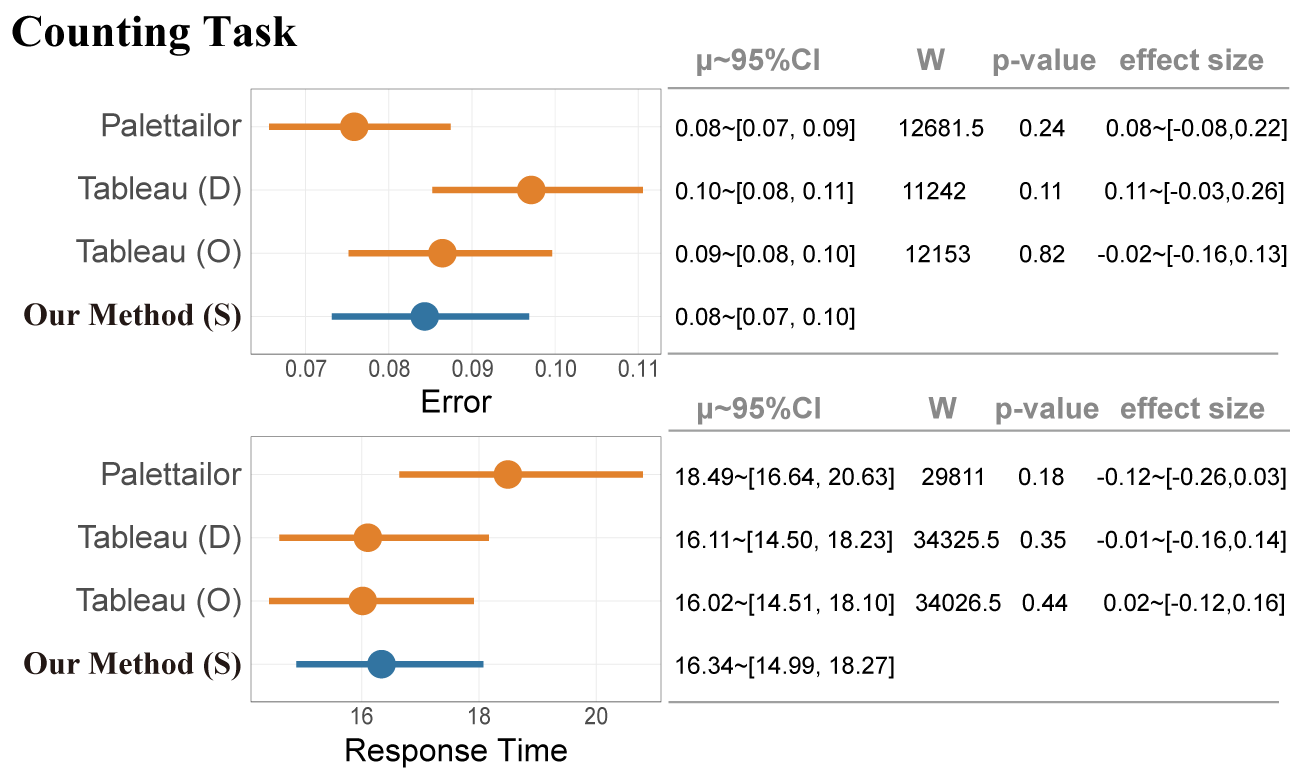}
\caption{Confidence interval plots and statistical tables for the \revised{counting task}. Error bars represent 95\% confidence intervals. Each table shows the statistical test results of our experimental condition (\emph{Our Method (S)}
with the three benchmark conditions (\emph{Palettailor}, \emph{Tableau (D)} and \emph{Tableau (O)}), showing the mean with 95\% confidence interval ($\mu \sim$ 95\%CI), W-value and p-value from the Mann-Whitney test, as well as  effect size (d $\sim$ 95\%CI).
}
\vspace*{-3mm}
\label{fig:countingTask}
\end{figure}

\vspace{.3em}
\noindent{\textbf{Results. }}
\revised{
Fig.\ref{fig:countingTask} shows the results of the visual discriminability experiment.
While \emph{Palettailor} achieves the best performance, our method performs better than the two \emph{Tableau} conditions. In particular, \emph{Tableau with default assignment} exhibited the worst performance. That said, there is no significant difference between these four conditions, implying a statistically similar performance. 
In terms of response time, we found that \emph{Tableau with optimal assignment} and \emph{Tableau with default assignment} take less time than our method and \emph{Palettailor}. However, again these differences were not statistically significant. The results overall indicate that our palette generation method is comparable to the benchmarks for the \emph{counting task} (\textbf{H1} confirmed).

We did not find a significant interaction between \emph{colorization methods} and \emph{cluster number} ($F(3, 1432) = 0.1342; p>0.1$). The effectiveness of the different methods on visual discriminability seems insensitive to the number of clusters.}

\subsection{Experiment 2: Interactive Exploration}
\label{subsec:Exp2}
We designed three tasks to examine the efficiency of our method for interactive exploration: a \emph{highlighting task} \revised{for measuring the emphasis effectiveness, and two tasks for measuring the context-preserving performance: a \emph{matching task} and a \emph{selecting task}}.

\vspace{.3em}
\noindent{\textbf{Conditions.}}
We included five conditions, the illustrations for different conditions can be found in the supplementary materials:
\begin{enumerate}
     \item \emph{Palettailor with lightness adjustment:}
        This condition represents a common highlighting strategy: applying lightness adjustments to a given colorized scatterplot that has good class discriminability to begin with. We maintain the original lightness level of the emphasized class while adjusting the lightness of all other classes. The adjusted lightness value depends on the background color. For example, if the background is white, the lightness should be high.

     \revised{\item \emph{Palettailor with alpha blending:}
        This condition represents another highlighting strategy: applying alpha blending to a given colorized scatterplot that has a good class discriminability to begin with. We set the opacity of the class to be emphasized to $1.0$ while adjusting the opacity of all other classes to $0.2$, which is recommended by Bartram et al.~\cite{bartram2010whisper}.

     \item \emph{Tableau Highlighter with default assignment:}
        For each color in the Tableau-10 palette, we obtain its corresponding faint color for the non-highlighted classes from the Tableau Highlighter
        . We applied this strategy to the default assignment of the Tableau palette.
}
     \item \emph{Tableau Highlighter with optimal assignment:}
        Similar to the above, but with an optimal assignment of the Tableau palette.
        

     \item \emph{Our Method (interactive):}
        Combining colors from the two contrastive palettes (salient and faint colors).
\end{enumerate}

\vspace{.3em}
\noindent{\textbf{Experimental Design.}}
Similar to the first experiment, we used a \emph{within-subject} design: each participant completed all 5 conditions across 12 scatterplots \emph{with a randomly chosen class to be highlighted} (60 stimuli in total).
To avoid ordering effects, we randomly shuffled the display order of stimuli. For each stimulus, we additionally randomly rotated the scatterplot. We also included four engagement checks to ensure participants were paying attention.

\begin{figure}[htb]
\centering
\includegraphics[width=1\linewidth]{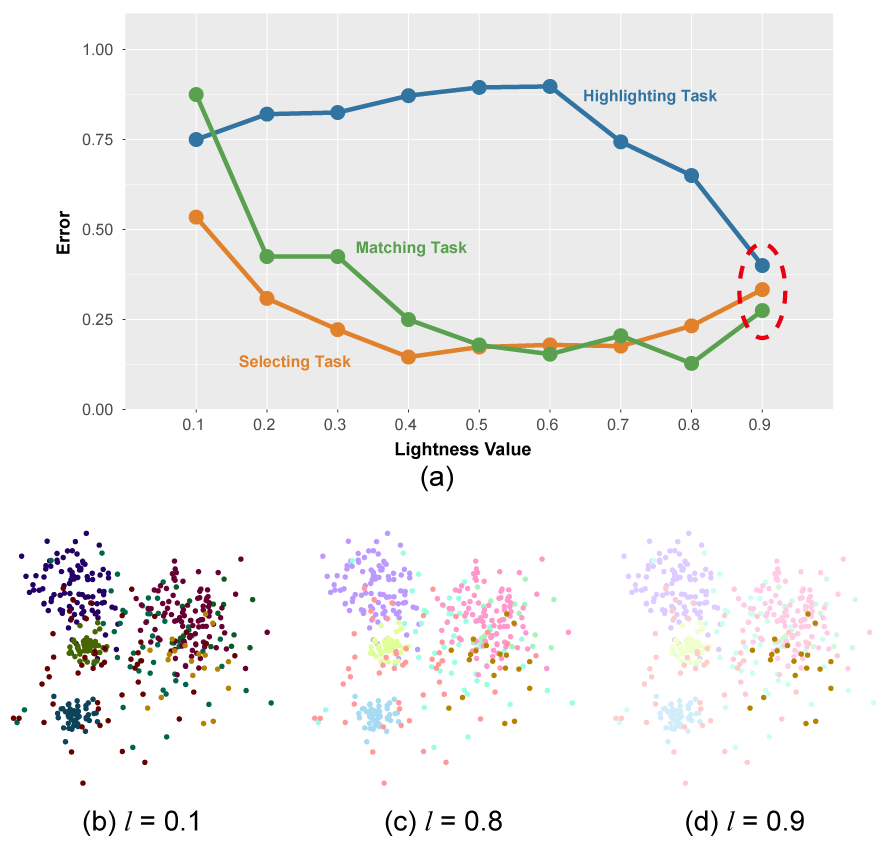}
\caption{(a) Results of the pilot study for selecting a proper lightness value: larger error value implies lower performance for each task. (b, c, d) Example trials used in the study with different lightness values. A lightness value of $0.9$ was selected as a sweet spot for the experiments.}
\vspace*{-3mm}
\label{fig:confirmLightness}
\end{figure}

\subsubsection{Pilot for Selecting Lightness Value}
\label{subsec:confirmLightnessValue}
~\\
One potential issue for using lightness to emphasize the desired class is that we cannot choose a value arbitrarily. We therefore conducted a pilot study across all three tasks, to determine an appropriate lightness level to assign to the non-highlighted classes.
We used four 8-class scatterplots in our pilot, and utilized \emph{Palettailor with lightness adjustment}. The lightness value varied incrementally within a range of [0.1, 0.9] and a step of 0.1. In total, we included 4 (scatterplots) $\times$ 9 (lightness levels) = 36 trials, plus 3 engagement checks. The trials were presented in random order. We recruited 10 participants for each task (30 participants in total) through AMT for a pilot. Participants who failed more than one engagement check were excluded, with new recruits taking their place, until we reached 10 participants. Each participant went through all 36 stimuli. All participants were US residents with a task-approval rate larger than 97\% and indicated normal color vision on their informed consent.

In Fig.~\ref{fig:confirmLightness}(a), we plot the average error rate for each lightness value of the three tasks: \emph{highlighting task}, \revised{\emph{selecting task}} and \revised{\emph{matching task}}. On a white background, error decreased for the highlighting task as the lightness value increased. This is due to the non-emphasized classes fading into the background, enabling the highlighted class to stand out, as shown in Figs.\ref{fig:confirmLightness}(b, c, d). Conversely, the errors for the \revised{selecting and matching} tasks improved with an increased lightness value. To reach the best performance among these tasks, we chose 0.9 as a reasonable lightness value for our experiment. 

\vspace{2mm}
\subsubsection{Highlighting task}
~\\
To evaluate whether our approach enables viewers to intuitively identify the emphasized class from a scatterplot, we conducted this task through AMT with 30 participants being accepted. The user interface is shown in Fig.\ref{fig:tasks}(c).
According to the completion time in the study (the details of the pilot study can be found in the supplementary materials), we paid each participant \revised{\$$1.00$} for the task based on the US minimum hourly wage.
No participant claimed color vision deficiency on their informed consent.


%


\begin{figure}[htb]
\centering
\includegraphics[width=1\linewidth]{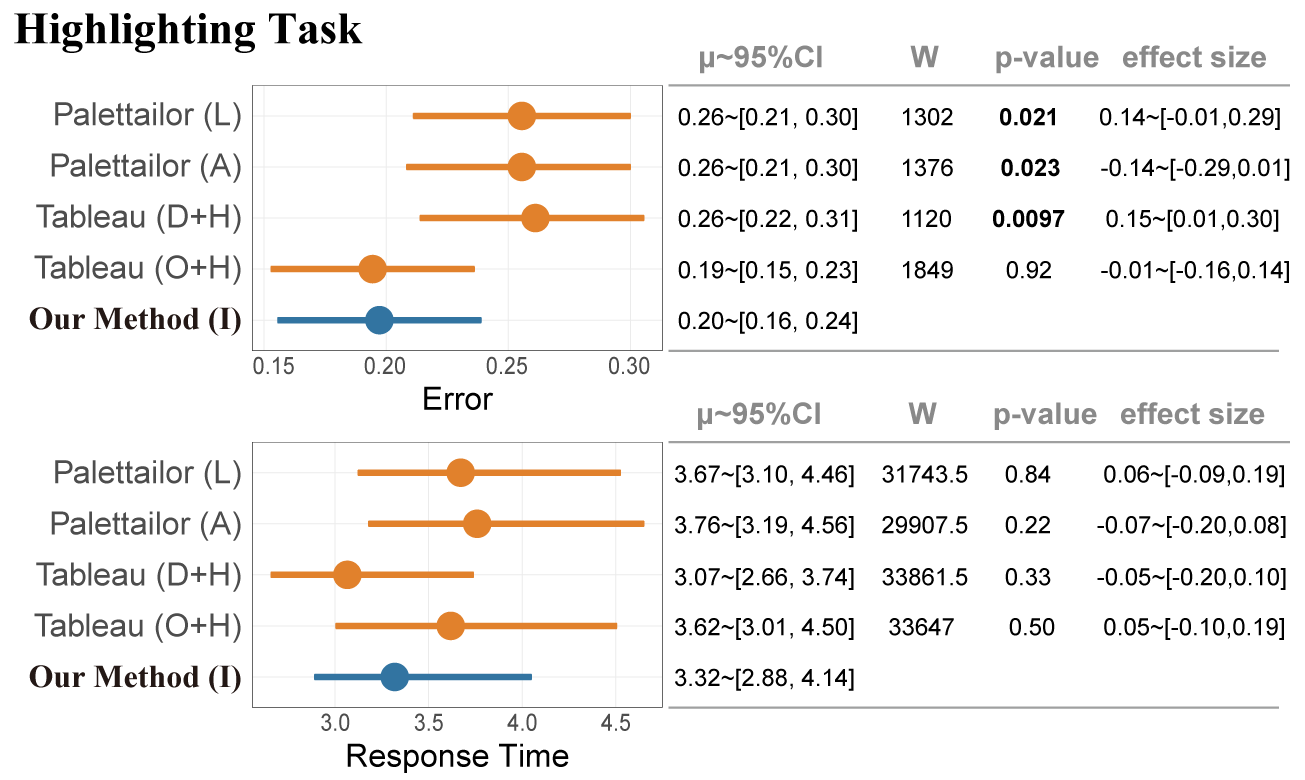}
\caption{Confidence interval plots and statistical tables for the highlighting task. Error bars represent 95\% confidence intervals. Each table shows the statistical test results of our experimental condition with the benchmark conditions (\emph{Palettailor (L)} indicates \emph{Palettailor with lightness adjustment}, \revised{\emph{Palettailor (A)} indicates \emph{Palettailor with alpha blending}, \emph{Tableau (D+H)} indicates \emph{Tableau Highlighter with default assignment}}, \emph{Tableau (O+H)} indicates \emph{Tableau Highlighter with optimal assignment}, \emph{Our Method (I)} indicates \emph{Our Method (interactive)}).
}
\vspace*{-3mm}
\label{fig:highlightingTask}
\end{figure}

\vspace{2mm}
\noindent{\textbf{Results. }}
\revised{
Fig.~\ref{fig:highlightingTask} shows the results of the  experiment for the \emph{highlighting task}.
\emph{Our Method} exhibited a significantly lower error rate than \emph{Palettailor with lightness adjustment} ($p=0.021$), \emph{Palettailor with alpha blending} ($p=0.023$), and \emph{Tableau Highlighter with default assignment} ($p=0.0097$). There was, however, no significant difference compared to \emph{Tableau Highlighter with optimal assignment} ($p=0.92$).
There were no significant differences in response time between our method and the benchmarks as the P value is more than 0.05 ($p>0.05$). We also did not find a significant interaction effect between the  \emph{colorization methods} and the \emph{number of clusters} ($F(4, 1790) = 0.2685; p>0.1$), meaning that visual emphasis is not affected by the number of clusters, which is consistent with the behavior of a popout effect.

The results indicate that our palette generation method outperforms commonly-used highlighting methods (e.g., lightness adjustment, alpha blending, and \emph{Tableau Highlighter with default assignment}), while being comparable to the best-case scenario of a state-of-the-art commercial system such as \emph{Tableau Highlighter with optimal assignment}. The results thus suggest an effective visual emphasis for our method, exceeding the performance of commonly applied manual highlighting techniques. We therefore consider \textbf{H2} to be confirmed.
}

\vspace{2mm}
\subsubsection{\revised{Matching task}}
~\\
As shown in Fig.\ref{fig:tasks}(d), we asked participants to select the cluster from the scatterplot whose color most closely matches the indicated color. The purpose of this color matching task was to examine whether our approach can maintain class recognition even when class color is changed in response to interactive highlighting.
We conducted this task through AMT with 30 participants being accepted.
According to the completion time in the study, we paid each participant \revised{\$$1.25$}.
No participant claimed color vision deficiency on their informed consent.

\begin{figure}[htb]
\centering
\includegraphics[width=1\linewidth]{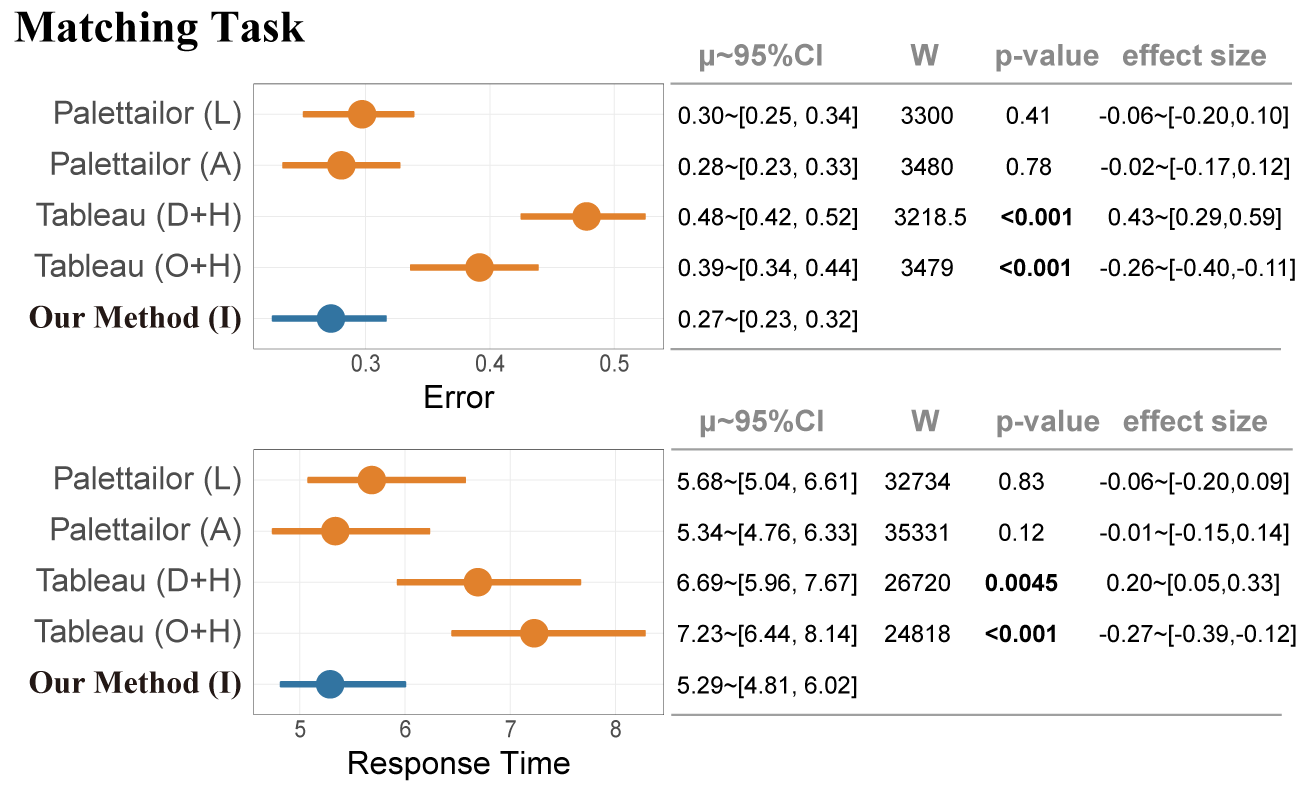}
\caption{
Confidence interval plots and statistical tables for the \revised{color-matching task}. 
}
\vspace*{-3mm}
\label{fig:constancyTask}
\end{figure}

\vspace{2mm}
\noindent{\textbf{Results. }}
\revised{
Fig.~\ref{fig:constancyTask} shows the results of the \revised{\emph{matching task}}.
Our method leads to a significantly lower error rate and response time compared to \emph{Tableau Highlighter with default assignment} and \emph{Tableau Highlighter with optimal assignment}, while it is slightly better than \emph{Palettailor with lightness adjustment} and \emph{Palettailor with alpha blending}. No significant interaction between \emph{colorization methods} and \emph{cluster number} was found ($F(4, 1790) = 2.163; p>0.05$).
The result indicates that our palette generation method has a better performance than the benchmark conditions for the \emph{matching task} w.r.t. color consistency, which confirms \textbf{H3}.

}
\vspace{2mm}
\subsubsection{\revised{Selecting task}}
~\\
We asked participants to select all colors from the palette that appear within a given circle. The user interface is shown in Fig.\ref{fig:tasks}(b). For this task, participants need to discriminate different colors around a small area, to examine how well the different methods can preserve the context of emphasized data. We conducted this task through AMT with 30 participants being accepted.
We paid each participant \revised{\$$2.00$} for an hourly wage consistent with the US minimum.
No participant claimed color vision deficiency on their informed consent.

\begin{figure}[htb]
\centering
\includegraphics[width=1\linewidth]{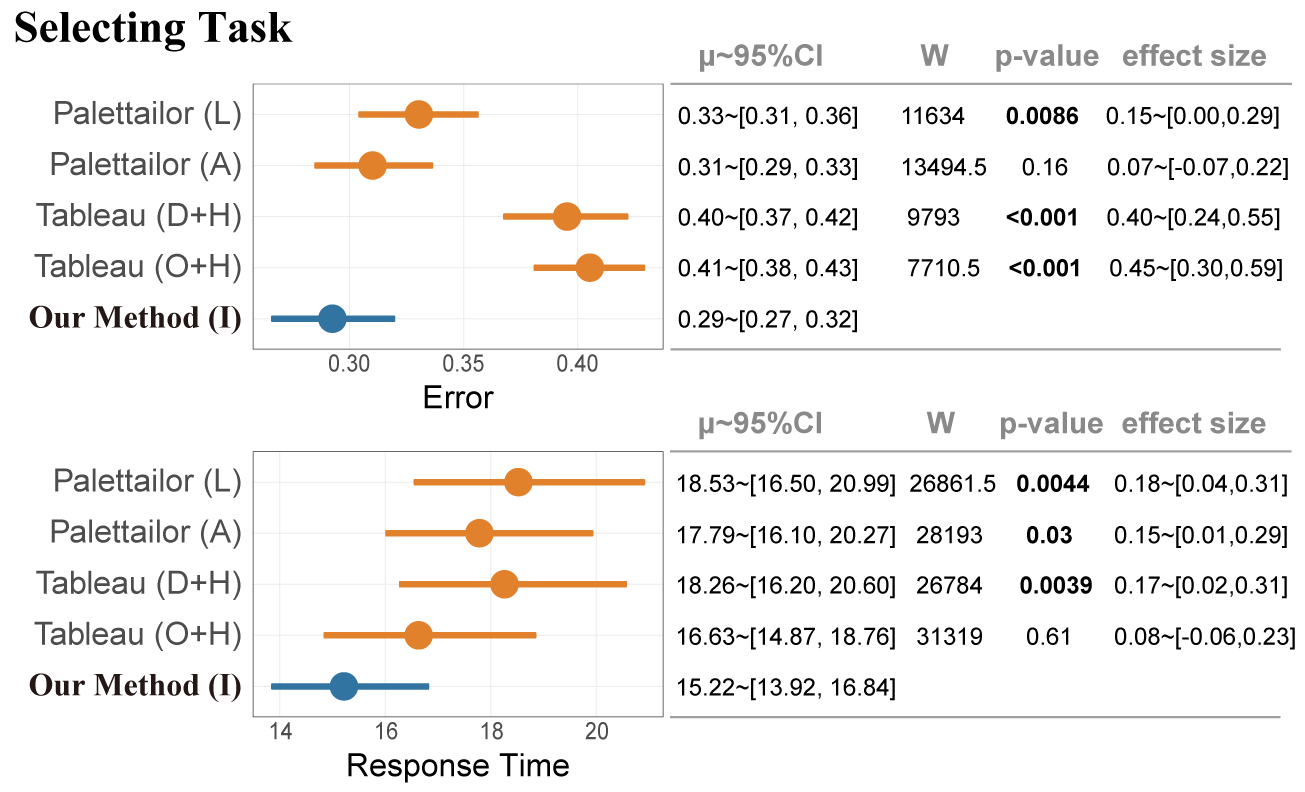}
\caption{
Confidence interval plots and statistical tables for the \revised{selecting task}. 
}
\vspace*{-3mm}
\label{fig:contextTask}
\end{figure}

\vspace{2mm}
\noindent{\textbf{Results. }}
\revised{
Fig.\ref{fig:contextTask} shows the results of the visual separability experiment for local discrimination.
Our method exhibits a significantly lower error rate relative to all other benchmark conditions, except \emph{Palettailor with alpha blending}. Although non-significant, we still achieved a better error rate than \emph{Palettailor with alpha blending} ($p=0.16$).
As for the completion time, our method achieves better performance than all other conditions, with a significantly shorter time than \emph{Palettailor with alpha blending} ($p=0.03$).
These results support \textbf{H3}. No interaction was found between \emph{colorization methods} and \emph{cluster number}  ($F(4, 1790) = 0.5798; p>0.1$).
}



\subsection{Discussion}
\label{subsec:discussionEval}

\begin{figure}[htb]
\centering
\includegraphics[width=1\linewidth]{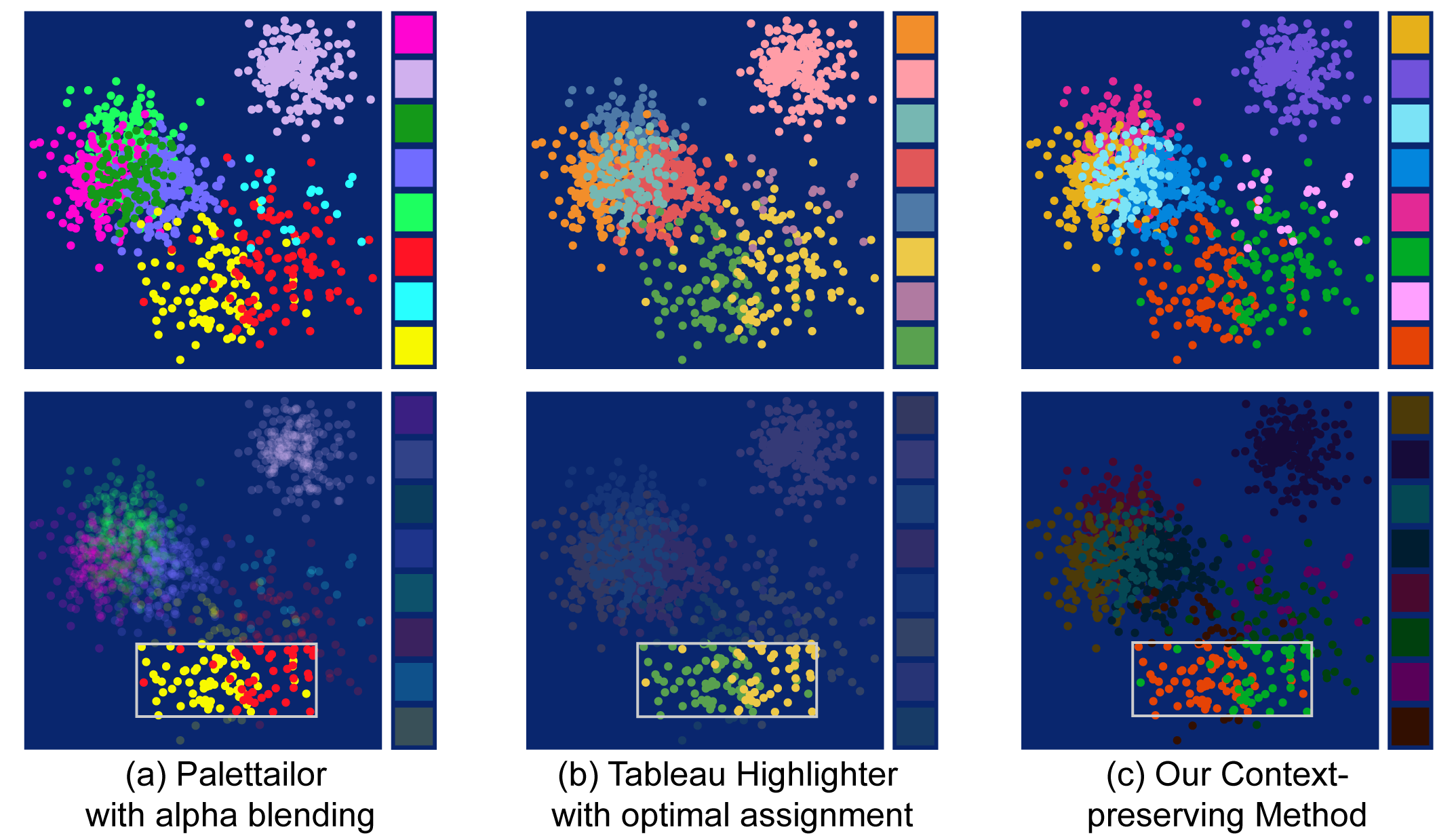}
\caption{
\revised{
Results generated for different methods with a blue background: \textbf{(a)} (top) palette and colorized scatterplot from Palettailor; (bottom) a highlighting effect is achieved by reducing the opacity of non-selected data points; \textbf{(b)} (top) Tableau palette and colorized scatterplot; (bottom) achieving a highlighting effect by applying Tableau Highlighter function; \textbf{(c)} (top)  salient palette and colorized scatterplot by our context-preserving highlighting method; (bottom) highlighting result by combining salient and faint color palette.}
}
\vspace*{-3mm}
\label{fig:background-tableau}
\end{figure}

We evaluated the effectiveness of our approach against the benchmark conditions through two crowdsourced experiments for two different scenarios (static visualization and interactive exploration).
\revised{
In the \emph{counting task} for a static visualization (see Fig.\ref{fig:countingTask}), we found that \emph{Palettailor} outperformed the \emph{Tableau} conditions and \emph{Our Method (static)}. This is reasonable since the design goal of \emph{Palettailor} is to maximize class discriminability. \emph{Our Method (static)} seems to be slightly better than \emph{Tableau with optimal assignment}. Notably, the latter achieves better performance than \emph{Tableau with default assignment}, which indicates that an optimal  assignment approach~\cite{Wang2018} does indeed improve discriminability for visualization. 
The results suggest that while \emph{Palettailor} outperforms our method in the \emph{counting task} for the global discriminability, the advantage is not substantial, thus representing a small overhead to pay for the ability to emphasize the desired classes.

For interactive exploration, our method shows a better performance. In the \emph{highlighting task}, we found that participants intuitively select the emphasized class in our approach. There is a significant advantage for \emph{Our Method} over some of the benchmark conditions (\emph{Palettailor with lightness adjustment}, \emph{Palettailor with alpha blending} and \emph{Tableau Highlighter with default assignment}). This indicates that our method attains better visual emphasis than most benchmarks while being comparable to the best-case scenario \emph{Tableau Highlighter with optimal assignment}). Interestingly, \emph{Palettailor with alpha blending} did not yield good highlighting performance. One reason is that colors from Palettailor might have a similar lightness to the background, e.g., light yellow class in Fig.\ref{fig:teaser}(b)-top. 
Another reason is that blended colors could inadvertently attract attention away from the desired class, some examples can be found in the supplementary materials.

As for the two context-preserving tasks, first, we found that in the color \emph{matching task}, \emph{Our Method (interactive)} performed better than \emph{Tableau Highlighter with default or with optimal assignment}, while achieving similar performance to \emph{Palettailor}, both with lightness adjustment and alpha blending. This is likely because our method, like other lightness adjustment approaches, works by only perturbing lightness while maintaining the original hue and saturation. 
%
When the background color is achromatic (white), during alpha blending, the hue will not be changed, thus achieving good performance.
However, for a chromatic background, alpha blending might result in poor class discriminability and color consistency (see Fig.~\ref{fig:background-tableau}). 
Since \emph{Our Method} also preserves name similarities for de-emphasized colors, it slightly outperforms \emph{Palettailor with lightness adjustment} and \emph{Palettailor with alpha blending}. An example illustration of this phenomenon can be found in the supplementary materials.
For the \emph{selecting task}, we found that our method achieves the best performance among all benchmark conditions, even though there was no significant difference to \emph{Palettailor with alpha blending}. However, \emph{Our Method} leads to a significantly shorter response time than the alpha blending approaches, likely because the latter   potentially introduces new blended colors that could distract the viewer.
}

\begin{figure*}[!ht]
   \centering
   \includegraphics[width=0.95\linewidth]{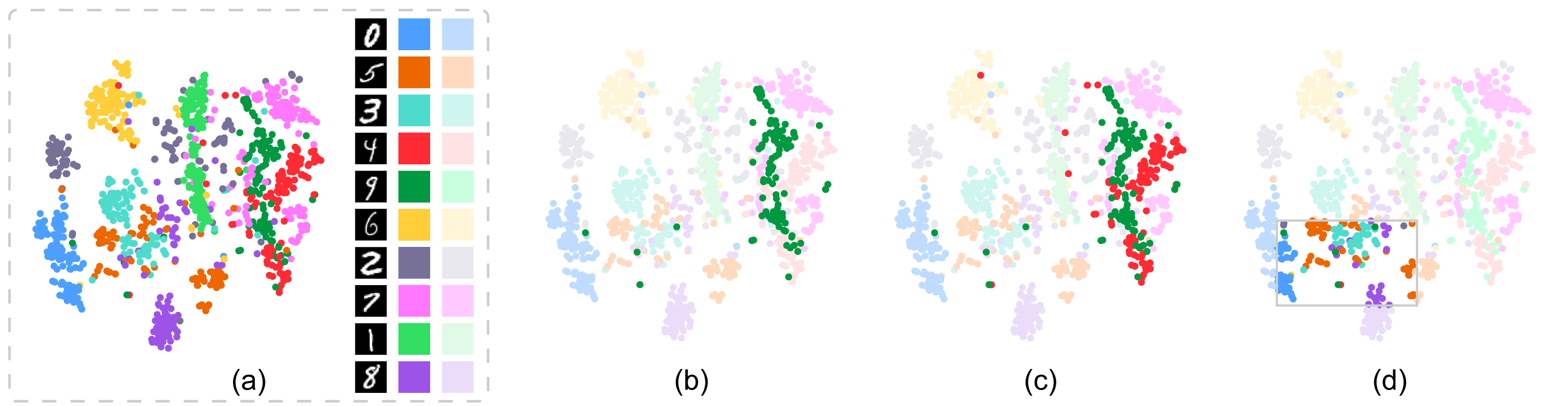}
   \caption{
   \revised{Exploring the MNIST dataset~\cite{lecun2010mnist} with our context-preserving highlighting technique. Result of static visualization (a) and the corresponding highlighting results by different selection methods: (b,c) legend selection, (d) brushing selection.}}
   \vspace*{-3mm}
   \label{fig:caseStudy}
\end{figure*}

The results indicate that our method maintains class discriminability for all classes while still achieving an intuitive highlighting effect. \revised{An added benefit to our context-preserving approach is that it automatically adapts to different backgrounds, thus producing more satisfactory results for chromatic backgrounds 
than Palettailor and Tableau Highlighter (see Fig.\ref{fig:background-tableau})}. 
A detailed analysis of response time, including the influence of class number, along with an analysis of potential speed-accuracy tradeoffs, can be found in the supplementary materials.

Our evaluation has some \textbf{limitations}:
First, we only tested two state-of-the-art colorization methods (\emph{Palettailor} and \emph{Tableau}, and their corresponding highlighting strategies). This choice was done to mitigate fatigue effects on participants. Whether other palettes (e.g., ColorBrewer's collection and Colorigorical) would lead to similar results remains to be seen.
Second, our experiment only focused on color-based highlighting; however, many methods exist using other visual variables to emphasize classes such as shape and mark size.
\revised{Third, the experimental setup is idealized: the scatterplots are relatively simple and the emphasis is applied to entire classes. The evaluation should therefore be extended with more complex datasets and tasks. We also did not measure participant preference (\eg from an aesthetics standpoint, where designer-crafted palettes might perform better than auto-generated results), leaving this aspect as future work.}
Finally, although we made attempts to reduce learning effects (e.g., random display order, randomly rotating scatterplots), some residual learning  could still have happened due to stimuli rotation. 

\section{System and Case Studies}
\label{sec:extension}
To aid designers in crafting categorical color palettes with contextual highlighting effects, we developed a web-based design tool that embodies our methodology\footnote{\small \url{https://palettailor.github.io/highlighting/}}.
Details of the system can be found in the supplementary materials. The interface allows users to select and highlight data via a variety of interactions, including clicking individual data points, clicking color legend to select an entire class, and brushing to select points that lie within a range. In the following, we present two extensions of our technique and  conducted two case studies on real-world  datasets.

\vspace{1.5mm}
\noindent\textbf{Extensions for Bar and Line Charts}.
In addition to scatterplots, our color mapping method can be easily extended to other categorical visualization types such as bar or line charts. This is achieved by treating each bar or line segment as a mark and then using the same method to compute their class contrasts, where the detailed description can be found in the supplementary materials.

\vspace{1.5mm}
\noindent\textbf{Extensions for Multi-view Visualizations}. 
\yh{Our technique can be extended to generate consistent color mapping schemes for multi-view visualizations of the same multi-dimensional data. For example, the line chart in Fig.~\ref{fig:case-line} displays trends of different classes, the bar chart shows the total number of each class. Following one of the multi-view consistency principles that the same nominal values in a field should be encoded by the same colors across different charts~\cite{qu2017keeping}, we generate the color mapping scheme for the view with most overlap between classes and apply this scheme to the other views. }

\vspace*{-1mm}
\subsection{Handwritten Digits Dataset}
\label{sec:caseStudy}
\revised{
Here, we analyzed the MNIST data of handwritten digits~\cite{lecun2010mnist}, which contains 784 data dimensions with ten classes. We project this dataset onto a 2D scatterplot using t-SNE with 1000 random distinct samples.
As shown in Fig.~\ref{fig:caseStudy}(a), our technique first colorizes the scatterplot with an overall good class discriminability. The user can click on the legend color to select the corresponding class -- in this case, the green class (see Fig.~\ref{fig:caseStudy}(b)), which represents the number $9$. She finds that this class is heavily overlapping with red, so she also clicks to select the latter (see Fig.~\ref{fig:caseStudy}(c)). She speculates that this might be caused by the similar appearance of the two numbers. To further investigate similar overlaps, she brushes over the scatterplot to select the left bottom region: the orange, sky blue, and purple classes representing $5$, $3$, and $8$, respectively.
During this exploration, our technique produces consistently good pop-out effects, as the emphasized data is interactively selected and de-selected (see Figs.~\ref{fig:caseStudy}(b, c, d)). Additionally, class separability and color consistency are well maintained regardless of which data subset is highlighted.
}

\vspace*{-1mm}
\subsection{Air Quality Dataset}
\label{sec:caseStudy2}
We conducted a second case study with a real-world dataset, this time using line and bar charts.
Here, we analyzed an air quality dataset provided by Vito et al.~\cite{DEVITO2008750} containing hourly recordings of a multi-sensor gas device deployed in an Italian city for two months in 2004. The dataset contains five classes corresponding to different gases: \emph{CO}, \emph{NMHC} (non-metanic hydrocarbons), \emph{$NO_x$} , \emph{$NO_2$} and \emph{$O_3$}.

\begin{figure*}[!ht]
    \centering
    \includegraphics[width=0.98\linewidth]{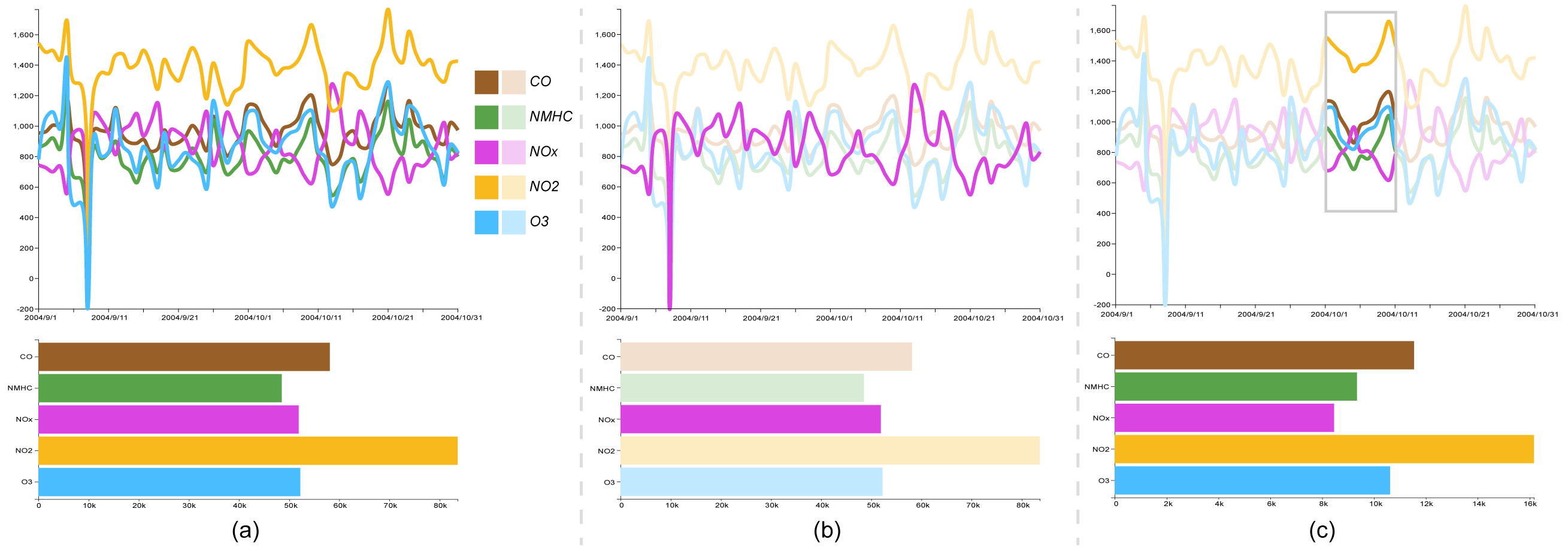}
    \caption{
    \revised{
    Visualizing an air quality dataset~\cite{DEVITO2008750} using two linked views with our context-preserving highlighting technique. \textbf{(a)} (top) colorizing the line chart by using a salient color palette; (bottom) applying the salient palette to a bar chart; \textbf{(b)} (top) highlighting result by combining salient and faint color mapping schemes with legend selection; (bottom) corresponding bar chart; \textbf{(c)} (top) highlighting result for the brushing selection; (bottom) corresponding bar chart for the selected data. Our method produces a good highlighting effect while maintaining class discriminability during interactive exploration.}}
    \vspace*{-3mm}
    \label{fig:case-line}
\end{figure*}

\revised{
Fig.~\ref{fig:case-line} shows line  and bar charts colorized using our technique, where each gas type is represented by a unique color. The line charts represent the gas change over time and the bar charts represent the total amount of each gas type.
We explore one class by interactively highlighting it through a legend selection. 
Fig.~\ref{fig:case-line}(b) emphasizes the pink class, which represents \emph{$NO_x$}. Our method achieves good overall class discriminability while allowing the user to still investigate any of the de-emphasized classes.
The brush selection results shown in Fig.~\ref{fig:case-line}(c), show that our technique maintains good separability between all trendlines, for both selected and non-selected classes.}
 This ability to interactively vary the highlight while still maintaining context makes our method especially suitable for interactive visual exploration.

\section {Conclusion and Future Work}
We presented a\revised{n interactive context-preserving color} highlighting approach for multi-class scatterplots. Our method allows viewers to  intuitively identify points of interest, while ensuring visual discriminability of all classes in a visualization, and maintaining a stable color mapping scheme during interactive exploration.
This goal is achieved by generating two contrastive palettes and then dynamically combining these two palettes, thus allowing for an interactively-variable focus effect. 
 In addition to modeling intra-class discriminability, our method also ensures sufficient contrast with the background.
We evaluated our approach through a crowd-sourcing study, which empirically demonstrates reliable highlighting and good class discrimination for our generated palettes. To help users generate such designs, we extended this method to other categorical visualizations such as bar charts and lines. In addition, we propose a web-based tool that implements our approach, enabling a quick, data-driven generation of palettes for a context-preserving emphasis effect.

\yh{Our user study focuses on contextual highlighting to points of interest in single view visualizations. In the future, we will investigate its effectiveness on tasks spanning multiple views (e.g., comparison tasks~\cite{Ondov19}).} 
In addition to color, other channels (e.g., shape~\cite{liu2021data} and mark size~\cite{smart2019measuring}) are known to have an effect on visual prominence, which could interact with our color-based highlighting approach. Future work could explore the possibility of modeling these factors to produce reliable intrinsic highlighting across multiple visual channels. 


Second, our approach produces colors that might not be friendly to people with color vision deficiency. 
Future work could thus extend our palette generation techniques to incorporate physiologically based models of color-vision deficiency~\cite{machado2009physiologically}. Such an extension could allow for color optimization with accessibility constraints. \revised{Aesthetic preferences should also be concerned in the automated colorization method to better serve users.}


Lastly, we evaluated the effectiveness of our palettes against a limited number of highlighting techniques. However, since there are many different highlighting methods, such as shape,  size, and animation, it would be interesting to fully investigate the strengths and limitations of these approaches for engendering a highlight effect.

\begin{acks}
This work is supported by the grants of the NSFC
 (62132017, 62141217), and Shandong Provincial Natural Science Foundation (ZR2022JQ32). The authors would like to thank Mi Feng, Michael Sedlmair, and Qiong Zeng for their fruitful discussion and support.
\end{acks}

\bibliographystyle{ACM-Reference-Format}
\bibliography{cosaliency}

\end{document}


\title{\revised{Interactive Context-Preserving Color Highlighting} for Multiclass Scatterplots}

\centerline{-- Supplementary Material --}
\vspace{5mm}

\author{Kecheng Lu}
\email{lukecheng0407@gmail.com}
\affiliation{%
  \institution{Shandong University}
  \country{China}
}

\author{Khairi Reda}
\email{redak@iu.edu}
\affiliation{%
  \institution{Indiana University-Purdue University Indianapolis}
  \country{United States}
}

\author{Oliver Deussen}
\email{oliver.deussen@uni-konstanz.de}
\affiliation{%
  \institution{University of Konstanz}
  \country{Germany}
}

\author{Yunhai Wang}
\authornote{corresponding author}
\email{cloudseawang@gmail.com}
\affiliation{%
  \institution{Shandong University}
  \country{China}
}
\renewcommand{\shortauthors}{Lu et al.}

\newcounter{partextdummy}
\newcommand*{\CreateLink}[2]{%
  \begingroup
    \renewcommand*{\thepartextdummy}{#2}%
    \ifhmode
      \raisebox{2ex}[0pt][0pt]{%
        \refstepcounter{partextdummy}%
        \label{#1}%
      }%
    \else
      \refstepcounter{partextdummy}%
      \label{#1}%
    \fi
  \endgroup
  \ignorespaces
}
\newcommand*{\LinkTo}{\ref}


\maketitle

This \textbf{supplementary material} provides additional experimental results for our submitted paper titled ``\revised{Interactive Context-Preserving Color Highlighting} for Multiclass Scatterplots''.

\paragraph{Navigation:}
\begin{enumerate}[start=1]
\item \LinkTo{probDist}
\item \LinkTo{alphaBlending}
\item \LinkTo{detailDiscussion}
\item \LinkTo{formalStudyDetails}
\item \LinkTo{interfaceInteraction}
\item \LinkTo{caseStudyScatterplot}
\item \LinkTo{caseStudyScatterplotMatrix}
\item \LinkTo{caseStudyLine}
\item \LinkTo{compensationDetails} 

\end{enumerate}


\newpage

\begin{figure*}[h]
\centering
\includegraphics[width=0.4\linewidth]{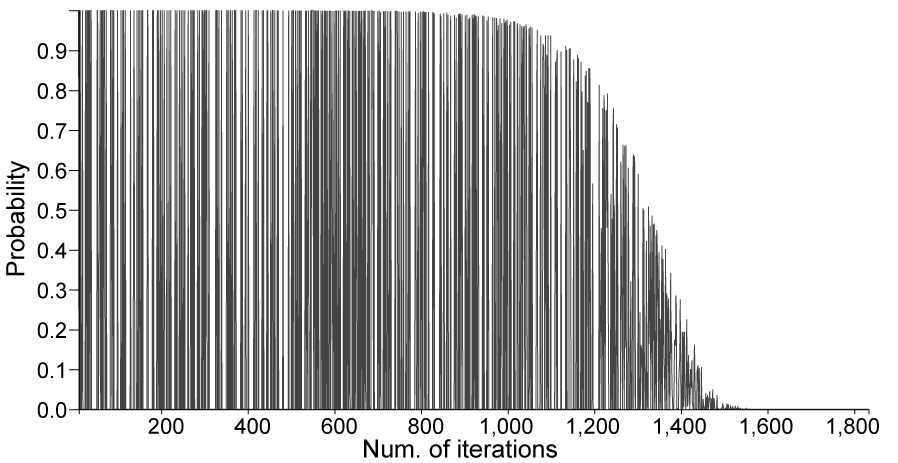}
\caption{The probability distribution of simulated annealing algorithm for finding the best lightness value.}
\label{fig:probDist}
\end{figure*}

\CreateLink{probDist}{Probability distribution for finding lightness value }
\paragraph{Probability distribution for finding lightness value}.
 To rapidly produce homogeneous backgrounds, we set a large probability for accepting a uniform lightness for all colors initially and decrease it as the number of iterations increases. As shown in Fig.\ref{fig:probDist}, at the beginning of the simulated annealing algorithm, the main process is to find the best uniform lightness for all colors. Then the probability is decreased according to the palette score and the number of iterations. Finally, we got the best lightness and mainly disturb the lightness of each color.

\begin{figure*}[h]
\centering
\includegraphics[width=0.98\linewidth]{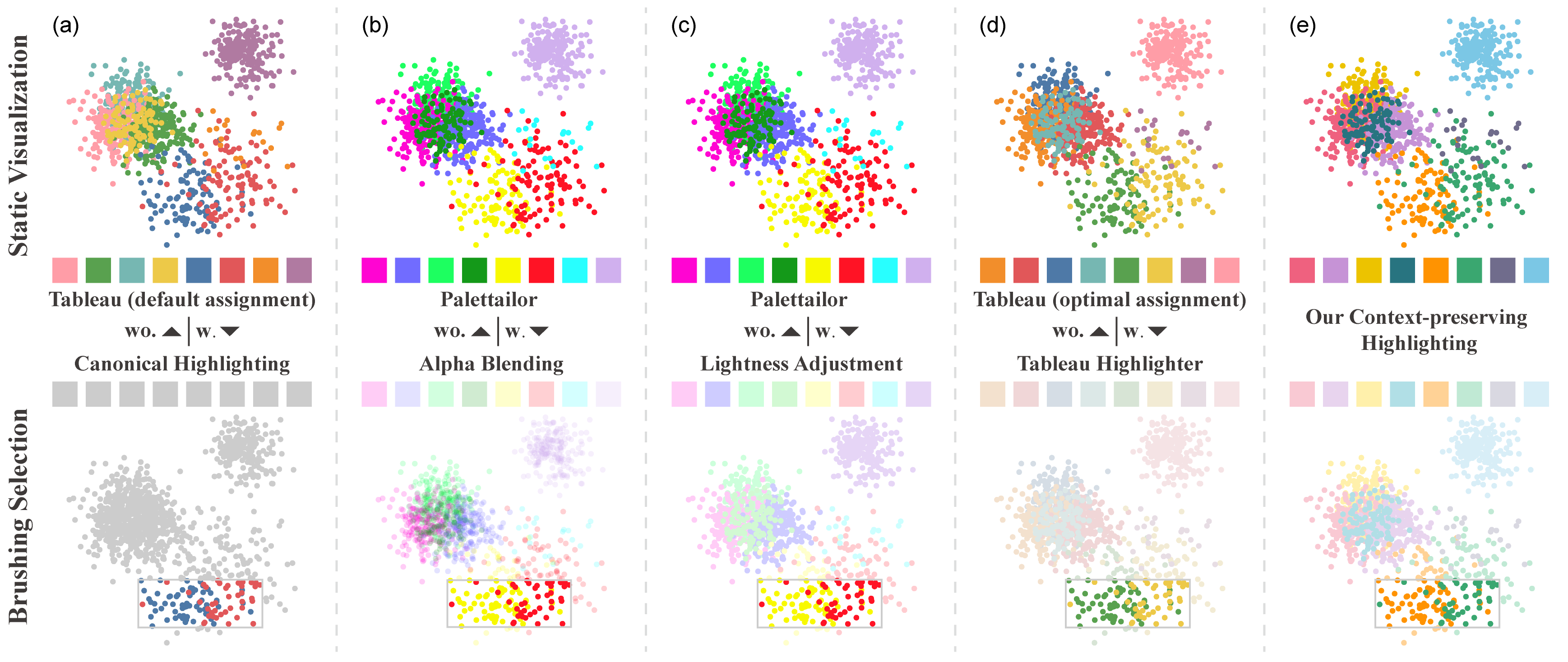}
\caption{Results for applying different color-based highlighting methods to brush a multi-class scatterplot. 
  \textbf{(a)} (top) result colorized by the Tableau palette and the default assignment; (bottom)
    a highlighting effect achieved by assigning a grey color to all non-selected data points;
  \revised{\textbf{(b)} (top) result colorized by a Palettailor-generated palette~\cite{Lu21};  (bottom) a highlighting effect is achieved by reducing the opacity of non-selected data points;}
  \textbf{(c)} (top) result colorized by a Palettailor-generated palette~\cite{Lu21};  (bottom) a highlighting effect is achieved by increasing the lightness of non-selected data points;
    \textbf{(d)} (top) result colorized by the Tableau palette and the optimal assignment; (bottom) achieving a highlighting effect by applying Tableau Highlighter function;
     \textbf{(e)} (top) result colorized by our method with the salient color palette; (bottom) our highlighting result by combining salient and faint color palettes.
     Our method allows highlighting a subset of data points while maintaining the discriminability of all non-selected points and color consistency of all pairs of color.}
\label{fig:teaser}
\end{figure*}

\CreateLink{alphaBlending}{Teaser with alpha blending}
\paragraph{Teaser with alpha blending}.
\revised{Due to the limited space of the paper, the full methods including \emph{Palettailor with alpha blending} are shown in this supplementary material. One straightforward way to preserve the context during highlighting is to modulate a visual factor (e.g., opacity) of non-selected data points. Yet, this method often leads to misleading colors in overlapping regions due to alpha blending, resulting in poor class separability (see the bottom in Fig.\ref{fig:teaser} (b)).}

\newpage
\CreateLink{detailDiscussion}{Detailed Discussion of the Results}
\paragraph{Detailed Discussion of the Results}.

\begin{figure}[!ht]
\centering
\includegraphics[width=0.98\linewidth]{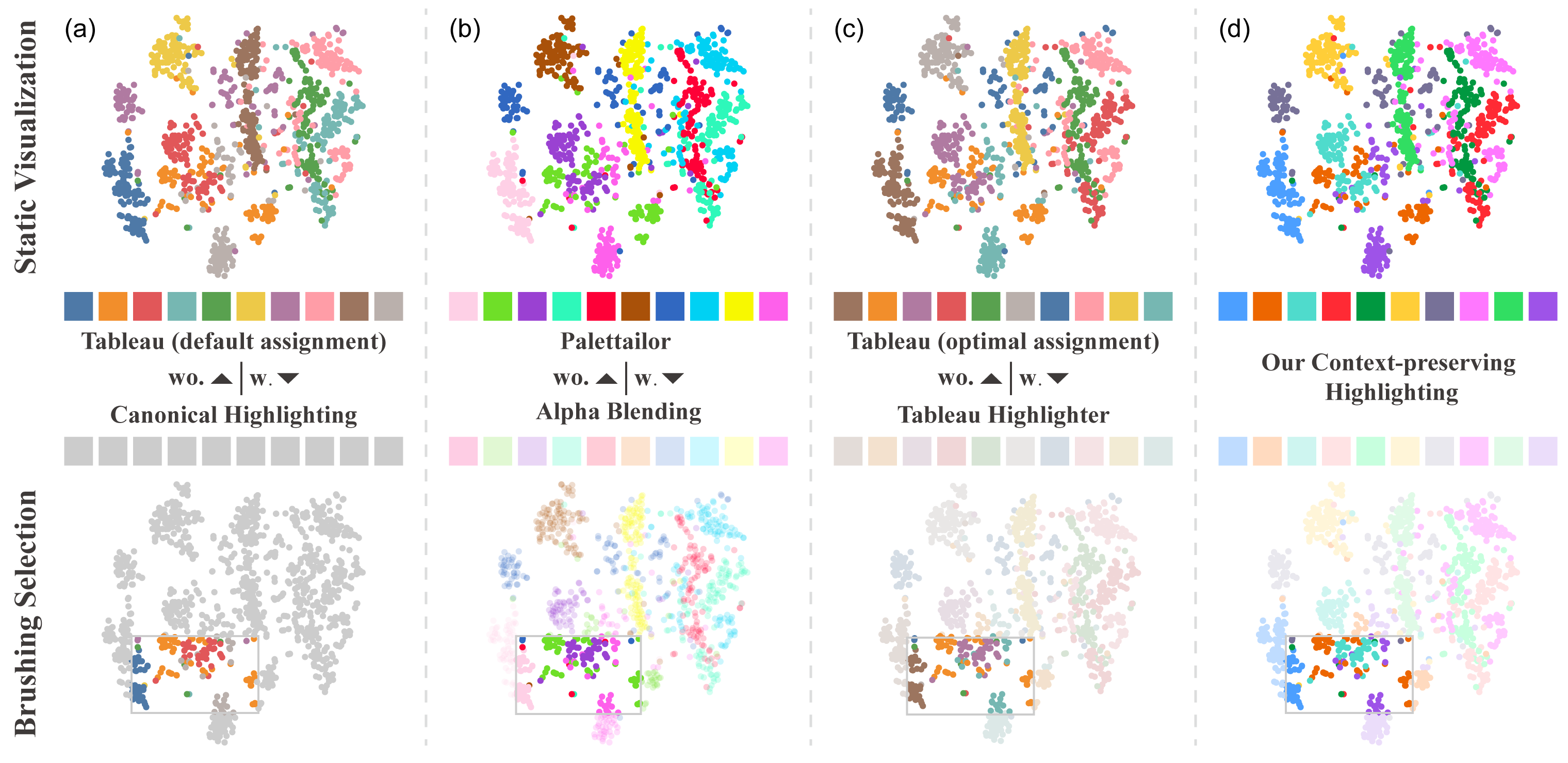}
\caption{
Visualizing the MNIST dataset~\cite{lecun2010mnist} with different methods for static visualizations (top row) and brushing selection (bottom row). \textbf{(a)} (top) Applying Tableau with default assignment to the data; (bottom) a highlighting effect achieved by assigning a grey color to all non-selected data points; \textbf{(b)} (top) using Palettailor to colorize the data; (bottom) applying alpha blending to highlight selected points; \textbf{(c)} (top) applying Tableau with optimal assignment; (bottom) using Tableau Highlighter to pop out the selected region; \textbf{(d)} (top) our method result for static visualization; (bottom) the corresponding highlighting result. Our method (d) gives a good highlighting effect while maintaining class discriminability during interactive exploration.}
\vspace*{-3mm}
\label{fig:caseStudy}
\end{figure}

We evaluated the effectiveness of our approach against the benchmark conditions through two crowdsourced experiments for two different scenarios (static visualization and interactive exploration).
\revised{
For the performance of the \emph{counting task} for static visualization, as shown in Fig.\ref{fig:trialsIllu}, we found that first, \emph{Palettailor} outperformed the two \emph{Tableau} conditions and \emph{Our Method (static)}. This is reasonable since the design goal of \emph{Palettailor} is to maximize class discriminability of a scatterplot while \emph{Tableau} is a designer-crafted palette for commonly used highlighting tasks and \emph{Our Method (static)} also serves such tasks. \emph{Tableau with optimized assignment} achieves better performance than \emph{Tableau with default assignment}, this indicates that the optimal discrimination assignment approach~\cite{Wang2018} improves the discriminability of the static visualization, as shown in Figs.\ref{fig:trialsIllu} (b, c). Specifically, \emph{Our Method (static)} seems to be slightly better than \emph{Tableau with optimal assignment}.
The results suggest that while \emph{Palettailor} beats our method in the \emph{counting task} for the global discriminability, the disadvantage for our method is not substantial, which represents a small overhead to pay for the ability to emphasize the desired class.

\begin{figure}[htb]
\centering
\includegraphics[width=1\linewidth]{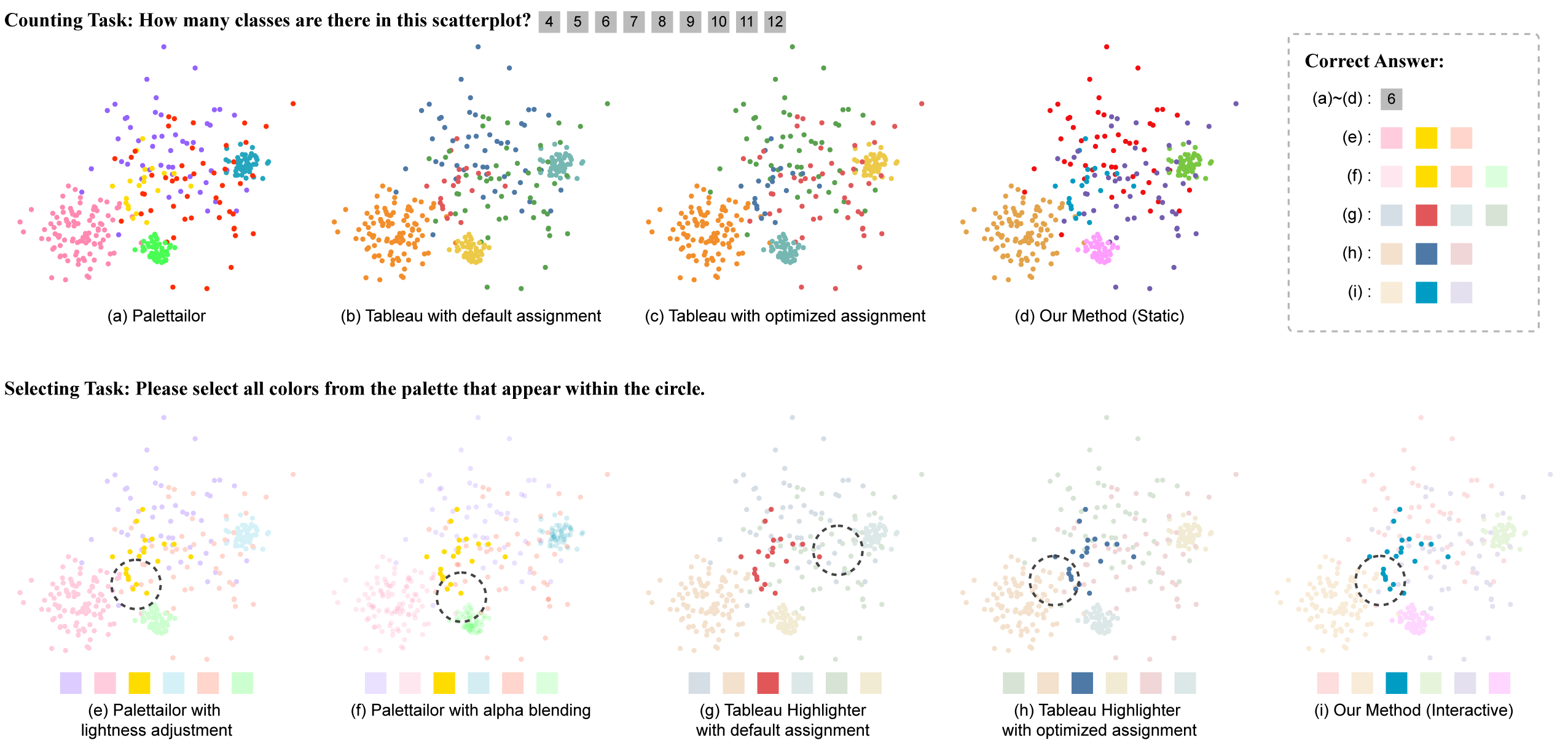}
\caption{
\revised{
One of the four six-class scatterplots used in the two experiments. There are four different colorization methods for the counting task in the top row: (a) Result generated from Palettailor; (b, c) Result generated by Tableau 10 palette with default and optimal assignment; (d) Our method result for static visualization. There are five different highlighting methods for the selecting task in the bottom row, the circles are randomly placed around the highlighted class: (e, f) using lightness adjustment and alpha blending to highlight the yellow class, the original palette is from (a); (g, h) using Tableau Highlighter to highlight the desired class based on (b) and (c), respectively; (i) Our method result for interactive visualization. The correct answer for each scatterplot is shown in the top right.}
}
\vspace*{-2mm}
\label{fig:trialsIllu}
\end{figure}

For interactive exploration, our method shows a better performance. In the \emph{highlighting task}, we found that without informing the participants what an emphasized class is, there's a significant difference between \emph{Our Method (interactive)} and some benchmark conditions (\emph{Palettailor with lightness adjustment}, \emph{Palettailor with alpha blending} and \emph{Tableau 
Highlighter with default assignment}). This indicates that, with regard to the common highlighting task, our contextual highlighting method performs better than standard highlighting methods.
\emph{Palettailor with alpha blending} did not get a good highlighting performance, one reason is that colors from Palettailor might have a similar lightness to the background, as shown in Fig.\ref{fig:trialsIllu} (f). The other reason is that the mixed color from alpha blending will be attractive, such as the red and blue classes shown in Fig.\ref{fig:caseStudy} (b).

We also found that \emph{Our Method (interactive)} has a similar performance with \emph{Tableau Highlighter with optimal assignment}, which implies that good discriminability helps the \emph{highlighting task} as well.
For the \emph{matching task}, \emph{Our Method (interactive)} performed better than \emph{Tableau Highlighter with default assignment} and \emph{Tableau Highlighter with optimal assignment}, while achieving similar performance to \emph{Palettailor with lightness adjustment} and \emph{Palettailor with alpha blending}. A possible explanation is that our method and lightness adjustment methods only perturb the lightness axis while maintaining hue and saturation. As for the alpha blending method, the major reason is that the background is white and the alpha blending does not change the color hue. An example can be found in Fig.\ref{fig:trialsIllu} (f), where the green and blue classes are similar to the original color in Fig.\ref{fig:trialsIllu} (a). However, given \emph{Our Method} that can maintain name similarity for the de-emphasized colors, it slightly outperforms \emph{Palettailor with lightness adjustment} and \emph{Palettailor with alpha blending}. An example illustration can be found in Fig.\ref{fig:trialsIllu} (e, f, i).
In the \emph{selecting task}, we found that \emph{Our Method (interactive)} achieves the best performance among all benchmark conditions, while there's no significant difference to \emph{Palettailor with alpha blending}. The explanation of this result is similar to the \emph{matching task}. However, \emph{Our Method (interactive)} takes a shorter time than the alpha blending method, since the alpha blending method blends the colors of overlapping marks, potentially introducing new colors, participants might be confused about these new colors, as shown in Fig.\ref{fig:trialsIllu} (f).
}

\newpage
\CreateLink{formalStudyDetails}{Class number analysis and speed-accuracy analysis}
\paragraph{Class number analysis and speed-accuracy analysis for each task of the formal study}.

\begin{figure*}[h]
\centering
\includegraphics[width=0.98\linewidth]{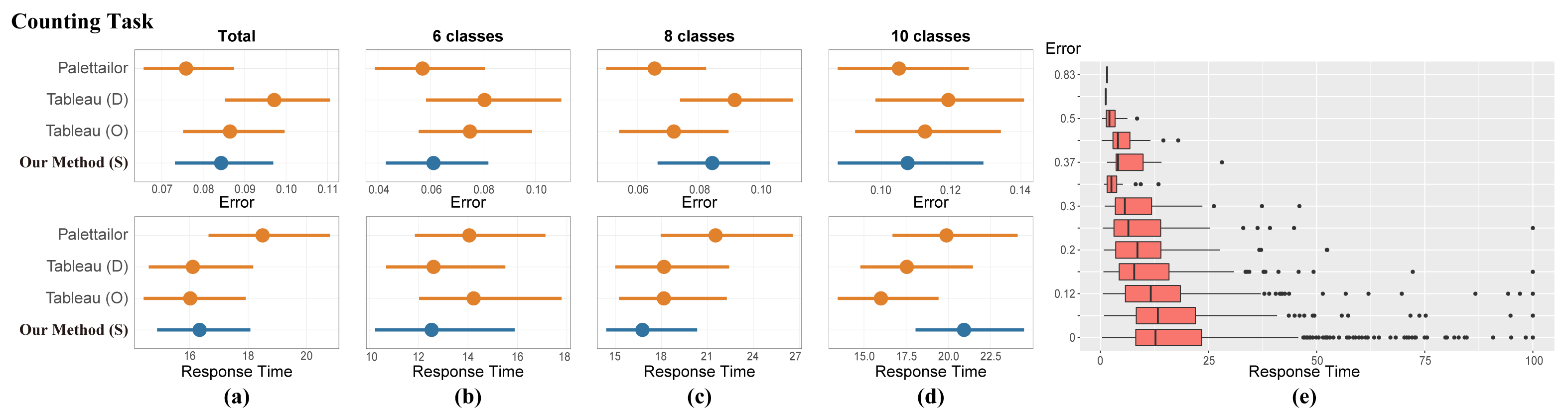}
\caption{
Statistics for the counting task. (a) Confidence interval plots for the whole trial data of different class numbers; (b) confidence interval plots for trial data which class number six; (c) confidence interval plots for trial data which class number is eight; (d) confidence interval plots for trial data which class number is ten; (e) box plots for the whole trial data of different class number.  Each table shows the statistical test results of our experimental condition (\emph{Our Method (S)}) with the three benchmark conditions (\emph{Palettailor}, \emph{Tableau (D)} and \emph{Tableau (O)}), showing the mean with 95\% confidence interval ($\mu \sim$ 95\%CI), W-value and p-value from the Mann-Whitney test, as well as  effect size (d $\sim$ 95\%CI).
}
\label{fig:countingTask}
\end{figure*}

\vspace{2mm}
\textbf{\emph{\revised{Counting task}}}.
To better understand how the different methods compare as the number of classes increases, we conducted a class number analysis for this class number counting task.
As shown in Figs.\ref{fig:countingTask}(a, b, c, d), we draw the confidence interval plots for the whole data and different class numbers. We can see that for different settings, the error rate and response time have similar performance. However, we found that in 10-class scatterplots, our method consumed more time than other methods while achieving less error rate, which gives an explanation for why \emph{Palettailor} always have a low error rate and high response time: users tend to spend more time to count how many classes are there in the scatterplot when the classes have a good separability. We did not find significant interaction effects between colorization methods and cluster number ( \revised{$F(3,1432) = 0.1342;p > 0.1$}). This means that the effectiveness of different methods on \emph{counting task} seems insensitive to the configuration of the cluster number.
We also provide a speed-accuracy analysis to show whether a speed-accuracy tradeoff exists. Since the relative error is categorical but not dichotomous (0/1), we conducted a Kruskal-Wallis test, where \revised{$Kruskal-Wallis chi-squared = 1134.2, df = 1130, p = 0.4596$}. The results indicate that the error and response time are weakly correlated. We also used some boxplots to inspect our data visually. If the data are weakly correlated, there will be a lot of overlap between the boxes. As shown in Fig.\ref{fig:countingTask}(e), the boxes are overlapped heavily.

\begin{figure*}[h]
\centering
\includegraphics[width=0.98\linewidth]{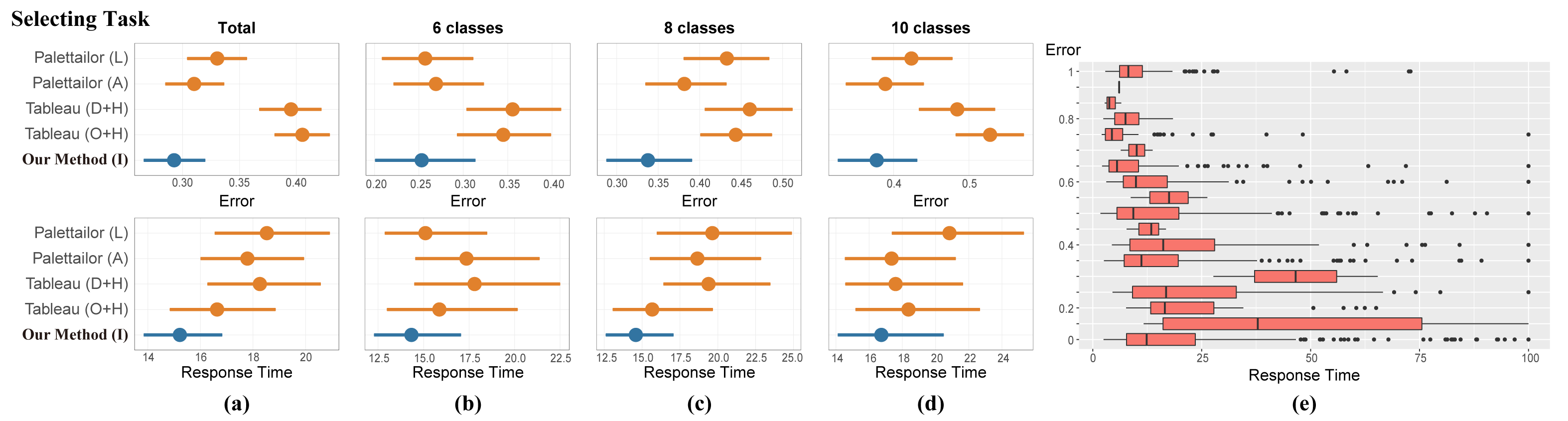}
\caption{Confidence interval plots and statistical tables for the selecting task. Error bars represent 95\% confidence intervals. Each table shows the statistical test results of our experimental condition with the benchmark conditions (\emph{Palettailor (L)} indicates \emph{Palettailor with lightness adjustment}, \emph{Palettailor (A)} indicates \emph{Palettailor with alpha blending}, \emph{Tableau (D+H)} indicates \emph{Tableau Highlighter with default assignment}, \emph{Tableau (O+H)} indicates \emph{Tableau Highlighter with optimal assignment}, \emph{Our Method (I)} indicates \emph{Our Method (interactive)}).
}
\label{fig:contextTask}
\end{figure*}

\vspace{2mm}
\textbf{\emph{\revised{Selecting task}}}.
The analysis is similar to the global discrimination task.
As shown in Figs.\ref{fig:contextTask}(a, b, c, d), we draw the confidence interval plots for the whole data and different class numbers. We can see that for different settings, the error rate and response time have similar performance. We did not find significant interaction effects between colorization methods and cluster number (\revised{$ F(4,1790) = 0.5798;p > 0.1$}). This means that the effectiveness of different methods on \emph{selecting task} for local discrimination seems insensitive to the configuration of the cluster number.
We also provide a speed-accuracy analysis to show whether a speed-accuracy tradeoff exists. Since the relative error is categorical but not dichotomous (0/1), we conducted a Kruskal-Wallis test, where \revised{$Kruskal-Wallis chi-squared = 1343.4, df = 1310, p = 0.2548$}. The results indicate that the error and response time are weakly correlated. We also used some boxplots to inspect our data visually. If the data are weakly correlated, there will be a lot of overlap between the boxes. As shown in Fig.\ref{fig:contextTask}(e), the boxes are overlapped heavily.

\begin{figure*}[h]
\centering
\includegraphics[width=0.98\linewidth]{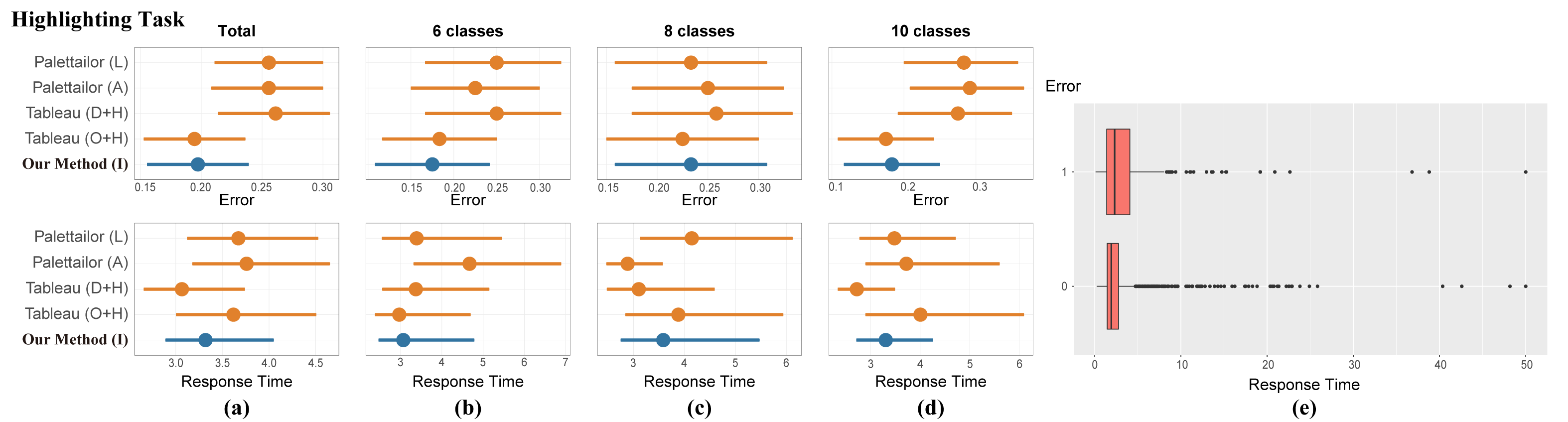}
\caption{Statistics for the highlighting task. Similar to the previous task.
}
\label{fig:highlightingTask}
\end{figure*}

\vspace{2mm}
\textbf{\emph{Highlighting task}}.
The analysis is similar to the previous tasks.
As shown in Figs.\ref{fig:highlightingTask}(a, b, c, d), we draw the confidence interval plots for the whole data and different class numbers. We can see that for different settings, the error rate and response time have similar performance. We did not find significant interaction effects between colorization methods and cluster number (\revised{$  F(4,1790) = 0.2685;p > 0.1$}). This means that the effectiveness of different methods on \emph{highlighting task} seems insensitive to the configuration of the cluster number.
We also provide a speed-accuracy analysis to show whether a speed-accuracy tradeoff exists.  Since the error of this task is dichotomous (0/1), we conducted a two-sample Wilcoxon rank sum test, where \revised{$W = 318223, p = 0.001926$}. The results indicate that the error and response time are strongly correlated. We also used some boxplots to inspect our data visually. If the data are strongly correlated, there will be a small overlap between the boxes. As shown in Fig.\ref{fig:highlightingTask}(e), the error(1) box has a larger response time than the correct(0) box. This is aligned with our experience: when the object is hard to find, it will take more time than an easier one.

\begin{figure*}[h]
\centering
\includegraphics[width=0.98\linewidth]{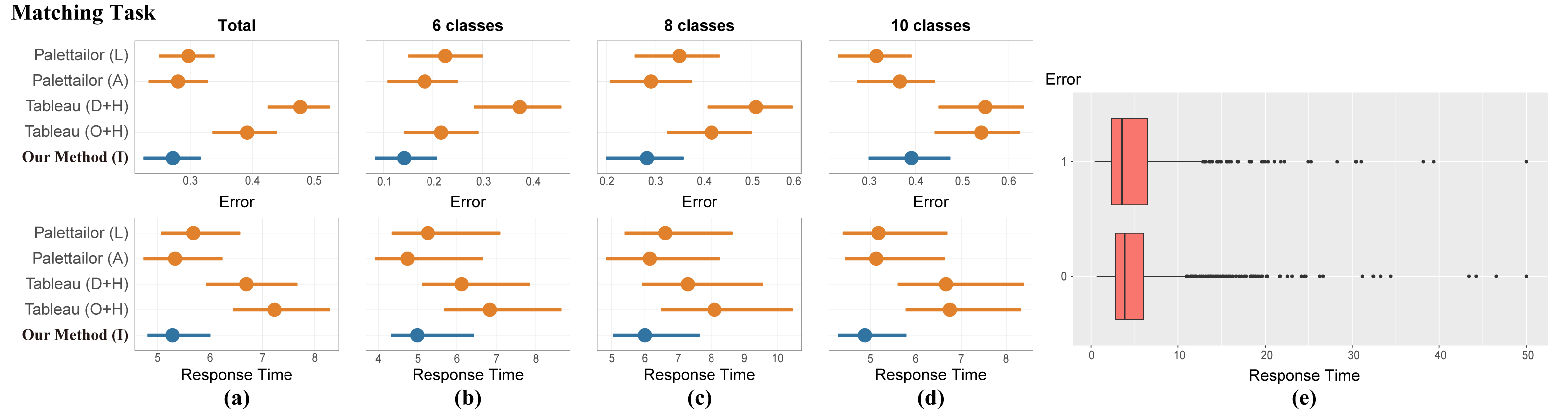}
\caption{Statistics for the color matching task. Similar to the previous task.
}
\label{fig:constancyTask}
\end{figure*}

\vspace{2mm}
\textbf{\emph{\revised{Matching task}}}.
The analysis is similar to the highlighting task.
As shown in Figs.\ref{fig:constancyTask}(a, b, c, d), we draw the confidence interval plots for the whole data and different class numbers. We can see that for different settings, the error rate and response time have similar performance. We did not find significant interaction effects between colorization methods and cluster number (\revised{$  F(4,1790) = 2.163;p > 0.05$}). This means that the effectiveness of different methods on \emph{constancy task} seems insensitive to the configuration of the cluster number.
We also provide a speed-accuracy analysis to show whether a speed-accuracy tradeoff exists.  Since the error of this task is dichotomous (0/1), we conducted a two-sample Wilcoxon rank sum test, where \revised{$W = 333316, p = 0.002$}. The results indicate that the error and response time are strongly correlated. We also used some boxplots to inspect our data visually. The boxes in Fig.\ref{fig:constancyTask}(e) have a large overlap, while the correct(0) box has a much larger number of outliers than the error(1) box. A possible explanation is that, as for the color constancy task, users often need to speculate the correct answer based on other colors, e.g., trials from \emph{Palettailor with lightness adjustment}, \emph{Palettailor with alpha blending} and \emph{Our Method(Interactive)} have better performance on this task, while some color still needs to be speculated from the context, results in more completion time but smaller error rate, while \emph{Tableau} methods will lead to colors that hard to be distinct and user might give up quickly, thus takes less time.

\newpage
\CreateLink{interfaceInteraction}{Interactive System and Extensions}
\paragraph{Interactive System and Extensions}.

\begin{figure}[ht]
	\centering
	\includegraphics[width=0.9\linewidth]{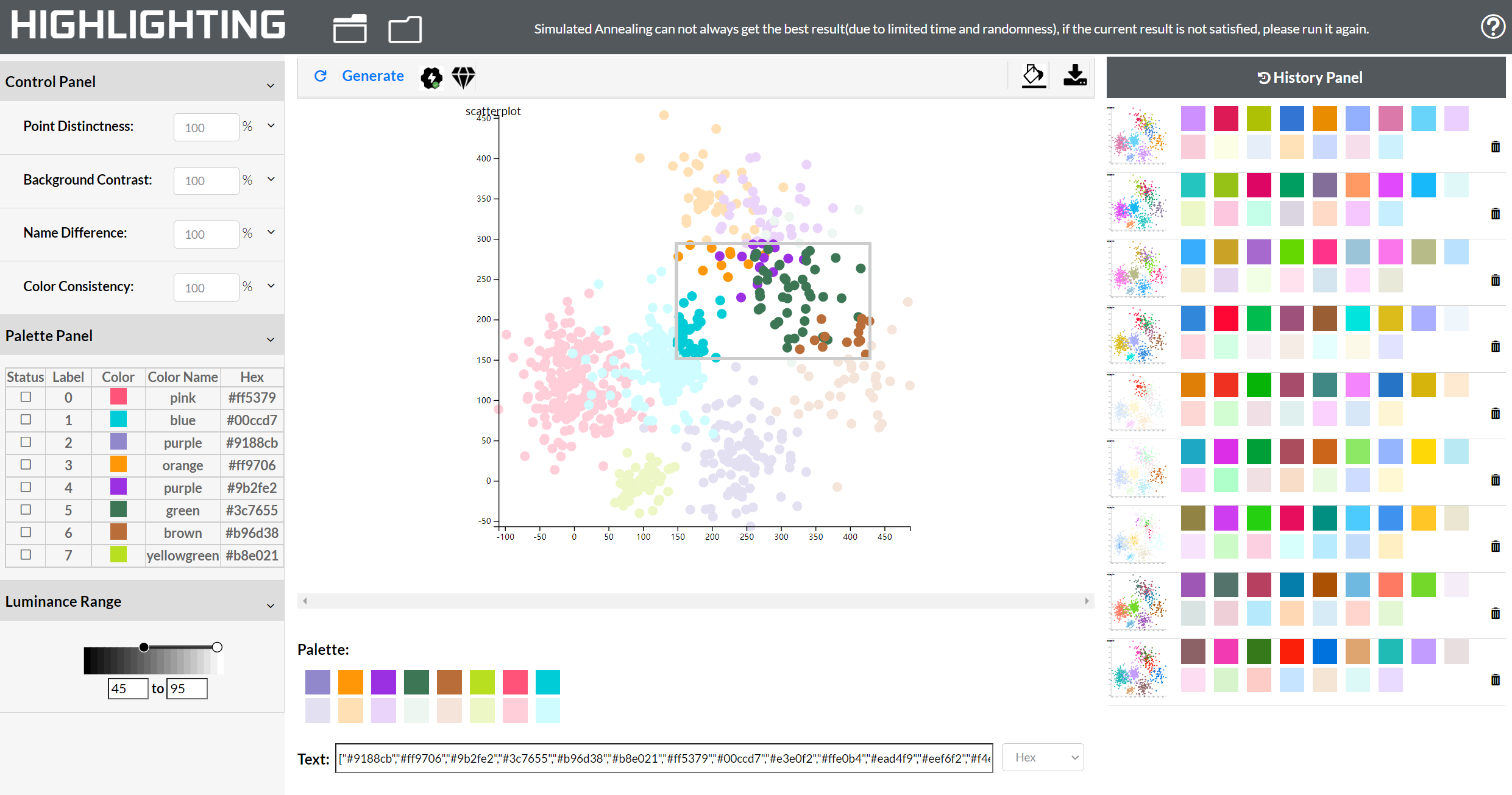}
	\caption{Screenshot of our interactive colorization system, which consists of four panels: (i) control panel; (ii) palette panel; (iii) visualization panel; and (iv) a history panel. }
	\vspace*{-3mm}
	\label{fig:ui-case}
\end{figure}

\begin{figure}[ht]
	\centering
	\includegraphics[width=0.9\linewidth]{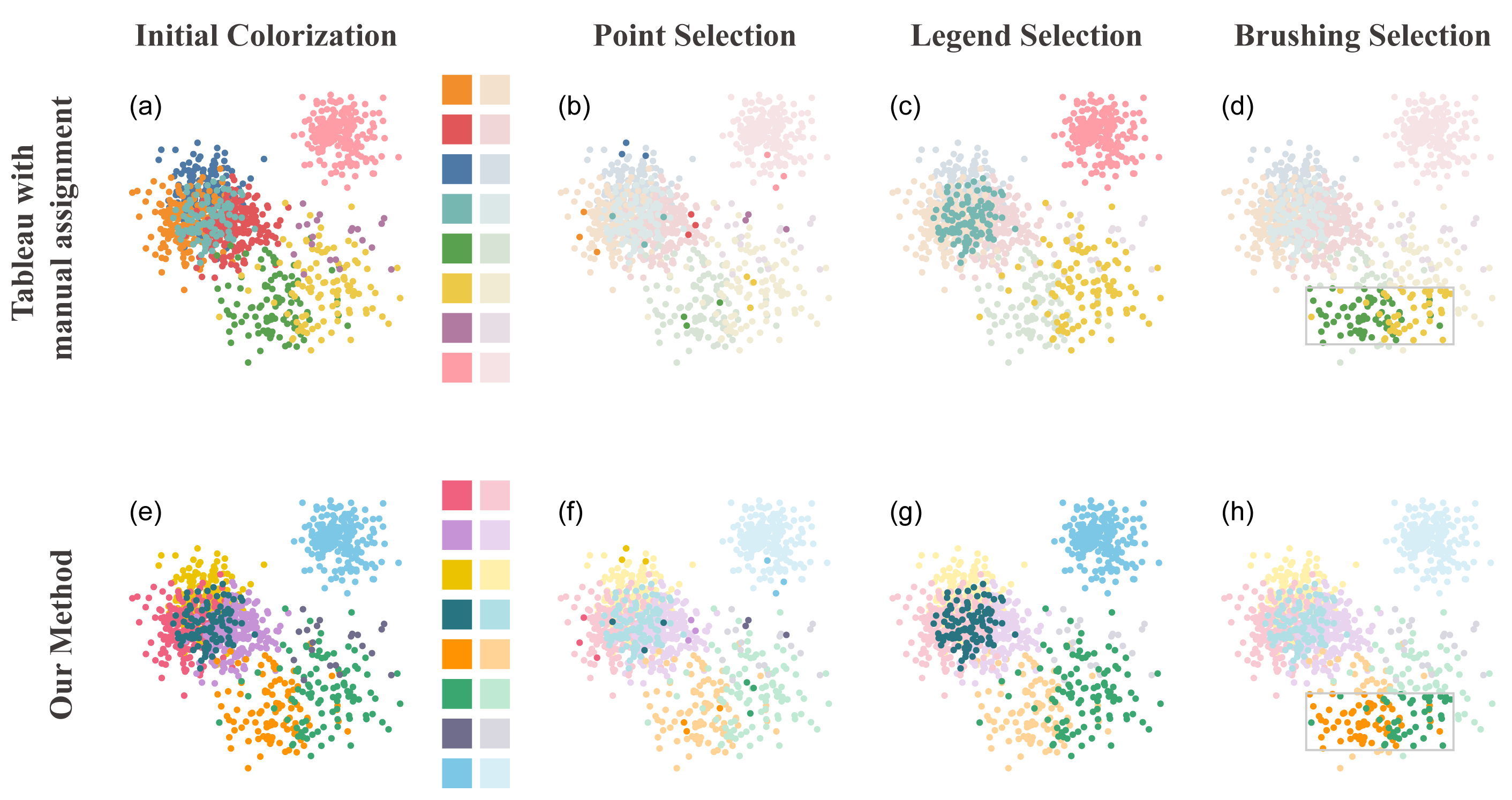}
	\caption{Different selections for the initial colorizations (a, e), including (b, f) point selection; (c, g) legend selection; (d, h) brushing selection. Our method achieves better or at least similar performance on different tasks than Tableau. }
	\vspace*{-3mm}
	\label{fig:supp-interaction}
\end{figure}

To aid designers in flexibly crafting categorical color palettes with contextual highlighting effects, we developed a web-based design tool that embodies our methodology\footnote{\small \url{https://anon-link.github.io/highlighting/}}.
The interface consists of four coordinated views: (i) a control panel, (ii) a palette panel for showing color information or adjusting the classes of interest, (iii) a visualization panel to provide a preview of the colorization result, and (iv) a history widget (see Fig.~\ref{fig:ui-case} for a screenshot). After uploading a labeled dataset, the system automatically finds an optimal color mapping scheme to colorize the input data. Classes are displayed on the palette panel, enabling the user to interactively highlight classes of interest. This adjustment forces the tool to automatically combine a new color palette from the pre-generated two palettes to emphasize the important classes. \revised{Besides, the user can directly select a subset of points from the visualization panel, using interactions like clicking or brushing, as shown in Fig.~\ref{fig:supp-interaction}.} The user can then save the resulting scheme using the history widget for future reference.

\newpage
\CreateLink{caseStudyScatterplot}{Case Study for Single Scatterplot}
\revised{
\paragraph{Case Study for Single Scatterplot}.
Here, we analyzed the MNIST database of handwritten digits provided by Yann et al.~\cite{lecun2010mnist}, which contains 784 data dimensions with ten classes. We project this dataset using tSNE onto a 2D scatterplot with 1000 random distinct samples. We first colorized the visualization using Tableau~\cite{tableau} (see Fig.~\ref{fig:caseStudy}(a)-top) with the default settings. To explore the details of the data distribution, the user often uses brushing to select interesting regions. Fig.~\ref{fig:caseStudy}(a)-bottom shows the brushing result from D3 or Vega-Lite, which simply applies a grey color to de-emphasized regions. There're two problems with this strategy: first, we lose all neighborhood information; second, the original grey color from the palette is similar to the de-emphasized grey color. The discriminability of the default assignment is not enough as well.
Then we tried the state-of-the-art automated colorization algorithm Palettailor~\cite{Lu21} (see Fig.~\ref{fig:caseStudy}(b)-top), and applied alpha blending to de-emphasize the surrounding regions, as shown in Fig.~\ref{fig:caseStudy}(b)-bottom, where the distribution of the light pink and yellow classes are hard to figure out and the light blue and cyan classes are mixed together.
Thanks to the optimal discrimination assignment approach~\cite{Wang2018}, we can improve the discriminability of the Tableau palette (see Fig.~\ref{fig:caseStudy}(c)-top), then we can use Tableau's Highlighter to interactively emphasize desired regions (see Fig.~\ref{fig:caseStudy}(c)-bottom). While the emphasis effect is good, the details of the other classes are difficult to distinguish, e.g., the pink class and red class are hard to distinguish.
In contrast to Tableau and Palettailor, our method is able to produce consistently good pop-out effects, with the emphasized region varied interactively as desired (see Fig.~\ref{fig:caseStudy}(d)). Additionally, class separability in our method is overall better than Tableau's Highlighter, e.g., the separability of pink and red classes in Fig.~\ref{fig:caseStudy}(d)-bottom is better than the pink and red classes in Fig.~\ref{fig:caseStudy}(c)-bottom. Uniquely, our method maintains good color consistency regardless of which class is being highlighted.
}

\newpage
\CreateLink{caseStudyScatterplotMatrix}{Case Study for Scatterplot Matrix}
\revised{
\paragraph{Case Study for Scatterplot Matrix}.
We conducted a second case study with a real-world dataset, this time using a scatterplot matrix.
Here, we analyzed a subset of FORCE 2020 Well well log and lithofacies dataset for Machine Learning competition~\cite{bormann_peter_2020_4351156}, which is used to predict lithology from existing labeled data using well log measurements. To simplify the visualization, we only show a handful of variables from the dataset, including \emph{RHOB}, \emph{GR}, \emph{NPHI}, \emph{DTC}, and \emph{LITH}. The first four of these five variables are numeric, and the last is categorical, which will be used as the class information.

\begin{figure}[!ht]
\centering
\includegraphics[width=0.98\linewidth]{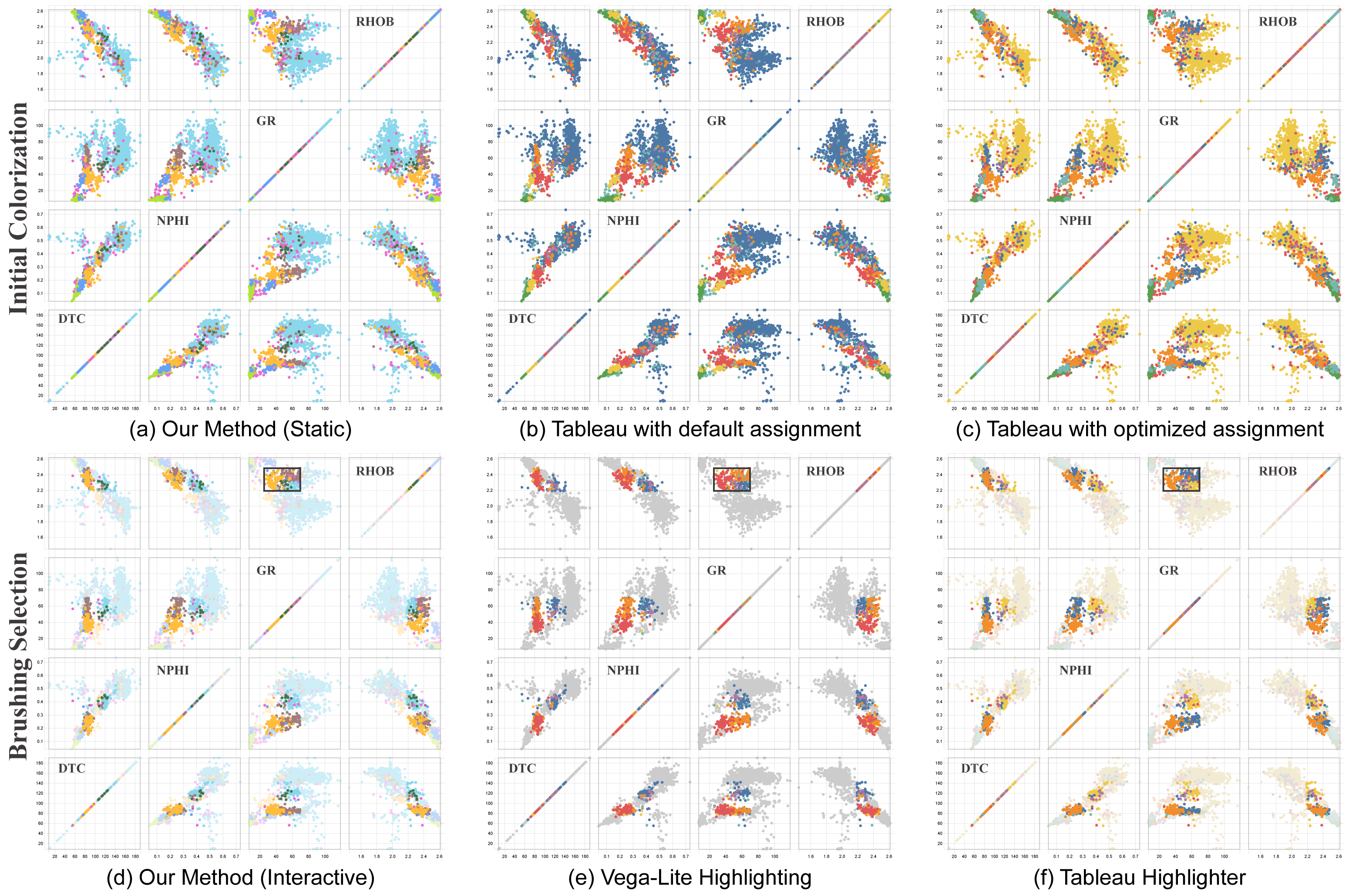}
\caption{
Visualizing the FORCE 2020 Well dataset~\cite{bormann_peter_2020_4351156} with different methods for static visualizations (top row) and brushing selection (bottom row).
}
\vspace*{-3mm}
\label{fig:caseStudy-matrix}
\end{figure}

Figs.~\ref{fig:caseStudy-matrix}(a, b, c) shows the scatterplot matrix colorized using Our Method(Static), Tableau with random assignment, and Tableau with optimal assignment, respectively. We can see that the optimal  assignment has better discriminability, but the classes in the corner are hard to be distinct, such as the green and light blue classes. Our method for static visualization achieves the best discriminability among these three results.
Figs.~\ref{fig:caseStudy-matrix}(d, e, f) shows the brushing selection results from the interactive exploration, the selected area is indicated by a black rectangle. We can see that Vega-Lite achieves the best highlighting but loses all the neighborhood information, while the emphasis effect of Tableau's Highlighter is good but the details of other classes are difficult to distinguish.
By comparison, the class separability of our method is better than Tableau for a baseline visualization (see Fig.~\ref{fig:caseStudy-matrix}(d).  With a focus on the interesting area, our method achieves better overall class discriminability than Tableau (Fig.~\ref{fig:caseStudy}(f)), allowing the user to still investigate any de-emphasized classes. The ability to interactively vary the highlight while still maintaining context makes our method especially suited for visual exploration.
}
\newpage
\CreateLink{interfaceInteraction}{Extensions for Bar and Line Charts.}
\paragraph{Extensions for Bar and Line Charts.}.
In addition to scatterplots, our color mapping method works also for other categorical visualization types such as bar or line charts. This is achieved by treating each bar or line segment as a point and then using the same method to compute their class contrasts.
This allows marks of interest to be highlighted while maintaining discriminability among all classes.
\begin{figure}[htb]
\centering
\includegraphics[width=0.9\linewidth]{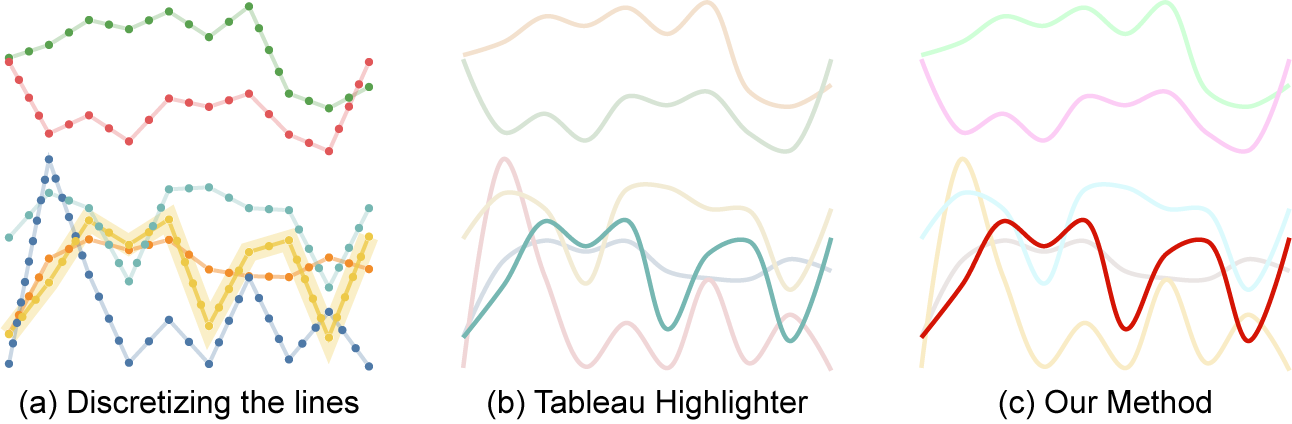}
\caption{
Extension for line charts in (a) with the yellow line to be highlighted. (a) Discretizing each line yields a point-based representation, which we take as input for our method; (b) result generated by using Tableau Highlighter; (c) result generated by our automated contextual highlighting method. Notice that higher discriminability between the non-highlighted classes in our method.
}
\vspace*{-2mm}
\label{fig:extensionCalc}
\end{figure}

Fig.~\ref{fig:extensionCalc} shows an example of colorizing a line chart. The highlighting effect of our automatic generation method in Fig.~\ref{fig:extensionCalc}(c) is not worse than an existing designer-crafted palette from Tableau-10 with Highlighter (see Fig.~\ref{fig:extensionCalc}(b)). Also here the discriminability between the other lines in the chart is maintained.

\CreateLink{caseStudyLine}{Case Study for Line Chart}
\paragraph{Case Study for Line Chart}.
We conducted a second case study with a real-world dataset, this time using line charts.
Here, we analyzed an air quality dataset provided by Vito et al.~\cite{DEVITO2008750} containing hourly recordings of a gas, multi-sensor device deployed in an Italian city from September 1 to October 31, 2004. The dataset contains five classes corresponding to different gases: \emph{CO}, \emph{NMHC} (non-metanic hydrocarbons), \emph{$NO_x$} , \emph{$NO_2$} and \emph{$O_3$}.

\begin{figure}[!ht]
\centering
\includegraphics[width=0.98\linewidth]{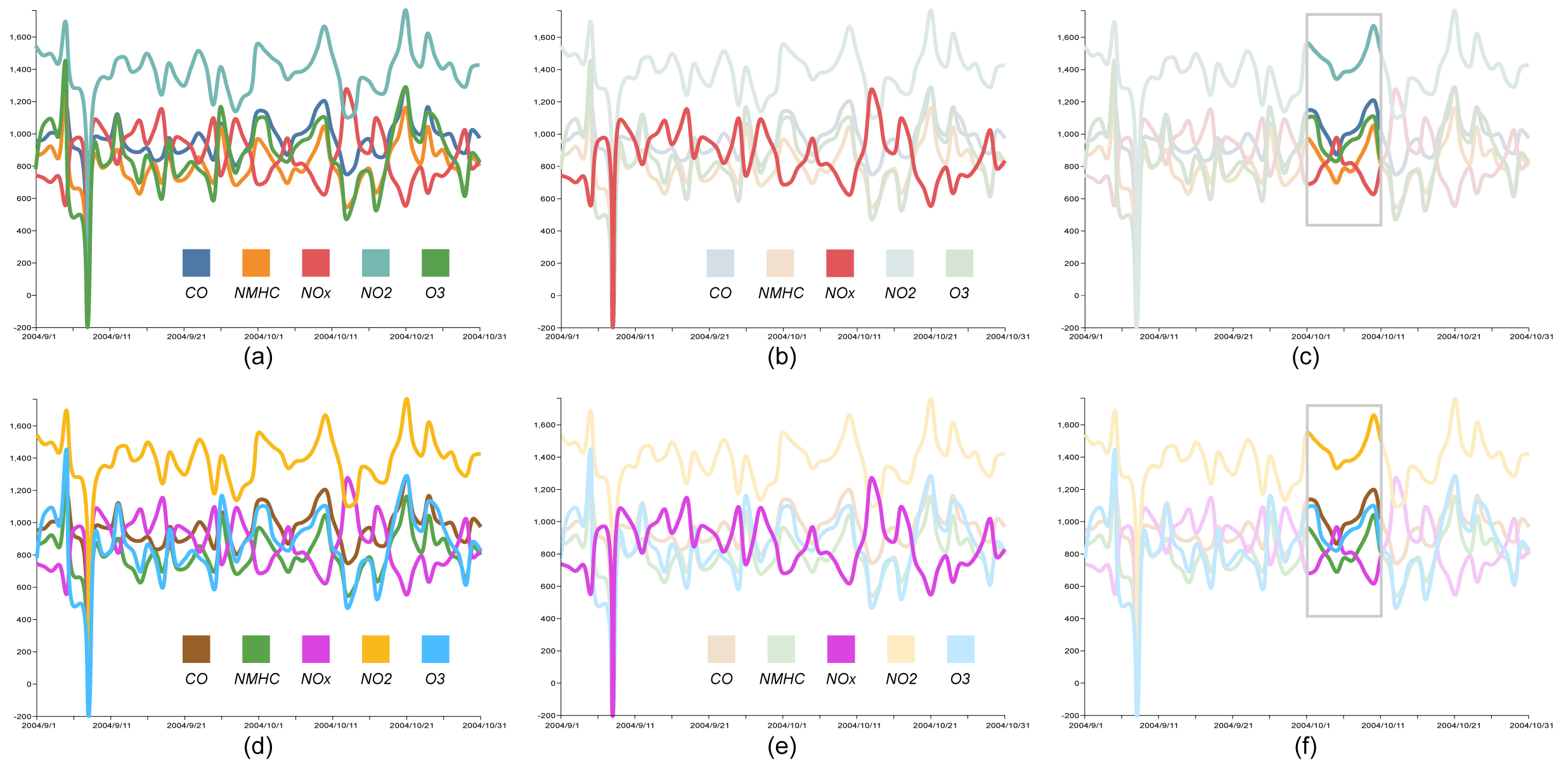}
\caption{
\revised{
Visualizing an air quality dataset~\cite{DEVITO2008750} with Tableau Highlighter (a) to highlight the trendline for \emph{$NO_x$} (b) and the brushing selection (c). The same purpose for our highlighting method (d, e, f). Our method (bottom row) gives a good highlighting effect while maintaining class discriminability during interactive exploration.}}
\vspace*{-3mm}
\label{fig:case-line}
\end{figure}

Figs.~\ref{fig:case-line}(a-f) shows line charts colorized using Tableau (top row) and our technique (bottom), where each gas type is represented using a unique color.
We explore one class by interactively emphasizing it using Tableau's Highlighter. Fig.~\ref{fig:case-line}(b) emphasizes the red class, which represents \emph{$NO_x$}. Here, the emphasis is good but the details of other classes are difficult to distinguish.
By comparison, the class separability of our technique is not worse than Tableau for a baseline visualization (see Fig.~\ref{fig:case-line}(d)).  With a focus on \emph{$NO_x$}, our method achieves better overall class discriminability than Tableau (Fig.~\ref{fig:caseStudy}(e)), allowing the user to still investigate any de-emphasized classes.
\revised{
From the brushing section results shown in Figs.~\ref{fig:case-line}(c, f), we can see that our technique maintains better separability between all data points while Tableau Highlighter results in a few similar colors for non-selected data points.}
 The ability to interactively vary the highlight while still maintaining context makes our method especially suited for visual exploration.

\newpage
\CreateLink{compensationDetails}{Compensation details}
\paragraph{Compensation details}.

The average spending time and the compensation for each task can be seen in Table.\ref{tab:summaryResults}.
We also provided the experiment data and analysis code in the supplementary materials. You can recreate all the experiment results shown in our paper through the source files (*.Rmd).

\begin{table}[htbp]
\centering
\Large
\caption{
 The average spending time and compensation for each task of the two crowd-sourcing experiments.}
\resizebox{0.8\linewidth}{!}{
\begin{tabular}{c|c|c}
 \hline
  Task & Average Spending Time & Compensation \\
 \hline
 \emph{Counting Task} & 13.87min & \$1.75\\

 \emph{Selecting Task} & 16.80min & \$2.00\\

 \emph{Highlighting Task} & 4.12min & \$1.00\\

 \emph{Matching Task} & 6.90min & \$1.25\\
 \hline
\end{tabular}
}
\label{tab:summaryResults}
\vspace{-1mm}
\end{table}

\bibliographystyle{../ACM-Reference-Format}
\bibliography{../cosaliency}